\documentclass[iop,apj,square,numbers,numberedappendix,appendixfloats]{emulateapj}
\usepackage[colorlinks, linkcolor=blue, citecolor=blue, urlcolor=blue]{hyperref}
\usepackage{amsmath,amssymb,bm,color}
\usepackage{graphicx}
\usepackage{epsfig}
\usepackage{color}
\usepackage{latexsym}
\usepackage{ulem}
\usepackage{url}
\usepackage{breakurl}

\shorttitle{Angular momentum evolution of stellar disks at high redshifts}
\shortauthors{Okamura et al.}

\begin{document} 

\title{Angular momentum evolution of stellar disks at high redshifts}

\author{Taku~Okamura\altaffilmark{1}, 
Kazuhiro~Shimasaku\altaffilmark{1,2},
and Ryota~Kawamata\altaffilmark{1}
}

\altaffiltext{}{Email: t.okamura@astron.s.u-tokyo.ac.jp}
\altaffiltext{1}{Department of Astronomy, Graduate School of Science, The University of Tokyo, 7-3-1 Hongo, Bunkyo-ku, Tokyo 113-0033, Japan}
\altaffiltext{2}{Research Center for the Early Universe, The University of Tokyo, 7-3-1 Hongo, Bunkyo-ku, Tokyo 113-0033, Japan}

\begin{abstract}
The stellar disk size of a galaxy depends on the ratio of the disk stellar mass to the halo mass, $m_\star \equiv M_\star/M_{\rm dh}$, and the fraction of the dark halo angular momentum transferred to 
the stellar disk, $j_\star \equiv J_\star/J_{\rm dh}$. Since $m_{\star}$ and  $j_{\star}$ are determined by many star-formation related processes, measuring $j_\star$ and $m_\star$ at various redshifts is essential to understand the formation history of disk galaxies.
We use the 3D-HST GOODS-S, COSMOS, and AEGIS imaging data and photo-$z$ catalog to examine $j_\star$ and $m_\star$ for star-forming galaxies at $z \sim$ 2, 3, and 4, when disks are actively forming.
We find that the $j_\star/m_\star$ ratio is $\simeq 0.77\pm 0.06$ for all three redshifts over the entire mass range examined, $8\times 10^{10} < M_{\rm dh}/h^{-1} M_\odot < 2\times 10^{12}$, with a possible ($<30\%$) decrease with mass. This high ratio is close to those of local disk galaxies, descendants of our galaxies in terms of $M_{\rm dh}$ growth, 
implying a nearly constant $j_\star/m_\star$ over past 12 Gyr. These results are remarkable because mechanisms controlling angular momentum transfer to disks such as inflows and feedbacks depend on both cosmic time and halo mass and indeed theoretical studies tend to predict $j_\star/m_\star$ changing with redshift and mass. It is found that recent theoretical galaxy formation simulations predict smaller $j_{\star}/m_{\star}$ than our values. We also find that a significant fraction of our galaxies appears to be unstable against bar formation. 
\end{abstract}


\section{Introduction} \label{sec1}
Within the $\Lambda$CDM paradigm, galaxies form in the center of hierarchically growing dark matter halos \citep{fall80}. In the tidal torque theory, gases and dark matter halos acquire angular momentum with log-normal distributions of the spin parameter through tidal gravitational fields 
 \citep{peebles69}. 
 The dimensionless spin parameter is given by 
\begin{eqnarray}
\lambda \equiv \frac{J |E|^{1/2}}{GM^{5/2}},
\end{eqnarray}
where $J$, $E$, and $M$ are the total angular momentum, total energy, and total mass of the system.
Since gases and halos share initial tidal torque fields, it is expected that gases and dark matter halos have the same amount of specific angular momentum. Gases gradually radiate away the thermal energy and then cool and collapse toward the center of dark matter halos. Their angular momentum halts the collapse and leads to a rotationally supported disk galaxy \citep{fall80, white91, mo98}. 
 
 In this formation scenario of disk galaxies, the disk size of a galaxy ($r_{\rm d}$) is given by  
 \begin{eqnarray}
 r_{\rm d} = \frac{1.678}{\sqrt{2}}\left(\frac{j_{\rm d}}{m_{\rm d}}\right)\lambda  r_{200} f_{\rm c}(c_{\rm vir})^{-1/2} f_{\rm R}(\lambda, c_{\rm vir}, m_{\rm d}, j_{\rm d}), \nonumber \\ \label{eq1_1}
 \end{eqnarray}
 \citep{mo98}.
 Here $j_{\rm d}/m_{\rm d}$ $( j_{\rm d} \equiv J_{\rm d}/J_{\rm dh},\ m_{\rm d} \equiv M_{\rm d} / M_{\rm dh}; \ {\rm d}:$star+gas) is the angular momentum retention factor and displays how much angular momentum acquired via tidal torques is conserved during the disk formation, $r_{200}$ is the radius of the dark matter halo within which encloses 200 times critical density, and $f_{\rm c}$ and $f_{\rm R}$ show, respectively, the difference in the density profile from an exponential profile and the gravitational effect of the disk. 
 By assuming that the angular momentum of a disk is fully conserved, $j_{\rm d}/m_{\rm d}=1$, this model successfully reproduces scaling relations of local disk galaxies: the stellar mass--size relation and the stellar mass--size scatter relation \citep{mo98, dutton12, fall12}.
 Because of the success of this picture, this model has been adopted in many semi-analytical models \citep[e.g.][]{somerville08,porter14,croton16}.  
 
However, the assumption that $j_{\rm d}/m_{\rm d}$ equals to unity independent of mass and cosmic time is not trivial, because highly-complex baryonic processes such as cooling, dynamical friction, and various feedback processes can change the specific angular momentum of disk galaxies. These processes are closely dependent on the mass of host dark matter halos. For example, the mass of dark matter halos controls how much expelled gases, which exchange the angular momentum with hot halo gases, can return to the galaxies again. The accumulation of such processes may increase or decrease the disk specific angular momentum. This is why the information of angular momentum is essential for comprehensive understanding of galaxy formation and evolution. It is important to understand the evolution of the angular momentum of galaxies as a function of dark halo mass at various redshifts.

In the present-day universe, since the pioneer work of \citet{fall83}, the angular momenta of galaxies with various morphological types and masses have been studied by observations and cosmological simulations \citep[e.g.][]{steinmetz99, governato07, fall12, fall13}. \citet{fall12} and \citet{fall13} have extended and updated the study of \citet{fall83} with recent observational data.
They have found that the specific angular momenta of spiral galaxies are not conserved, with $j_{\rm d}/ m_{\rm d} \simeq 0.6$  independent of halo mass. 
This implies that some baryonic processes mentioned above decrease the disk specific angular momentum.  Recent semi-analytical and hydrodynamical galaxy formation models have also obtained low angular momentum retention factors \citep{sales12,colin16, stevens16}. The roles of baryonic processes that determine the disk specific angular momentum have been examined: they include various types of feedback processes and the formation of bulges by disk instabilities.

On the other hand, beyond $z\sim1$, there are only a few studies that have observationally examined the specific angular momentum of galaxies because of the difficulty in obtaining kinematic measurements. \citet{burkert16} have analyzed the angular momenta of 359 disk star-forming galaxies at $z\sim 0.8-2.6$ and found $j_{\rm d}/m_{\rm d} \simeq 1$. \citet{contini16} have found in 28 low mass galaxies at $z\sim1$ almost the same stellar mass--angular momentum relation as the local one. However, some semi-analytical and hydrodynamical models predict that disk galaxies at $z\sim 1$ have smaller specific angular momenta than local galaxies \citep[e.g.][]{sales12, pedrosa15, stevens16}. Some results of cosmological galaxy formation simulations support the picture in which disk galaxies gradually acquire specific angular momentum as they grow. A consensus has not been reached on the angular momentum evolution beyond $z\sim1$. More observational data are needed to test the model predictions.

In this paper, to tackle the issue of the angular momentum evolution of disk galaxies and understand the formation and evolution of galaxy disks, we study the relation between the fraction of the dark-halo angular momentum transferred to the stellar disk ($j_{\star}:\ \star:$star) and the stellar to dark matter halo mass ratio ($m_{\star}$) at $z\sim$ 2, 3, and 4. We estimate dark halo masses by two independent methods: clustering analysis and abundance matching technique. In order to measure $j_{\star}$, it is popular to analyze galaxy kinematics with spectroscopy. However, it is very difficult to construct a large spectroscopic sample at high redshifts. Instead, we make use of the analytical model of \citet{mo98} that connects disk size with angular momentum. By measuring the disk sizes of galaxies and assuming this analytic model, we estimate $j_{\star}$.

\citet{kravtsov13} has investigated stellar disk size to halo size ratios $(r_{\rm d}/r_{\rm dh})$, which also reflect angular momentum retention factors, for local galaxies with a similar approach. \citet{kawamata15} and  \citet{shibuya15} have extended his study to high redshift galaxies and found that the disk size to halo size ratios are almost flat out to high redshift. Recently, \citet{huang17} and \citet{somerville17} have examined the disk size to halo size ratios as a function of stellar mass in more detail out to $z\sim3$ from CANDELS surveys using abundance matching. They have found that the disk sizes are proportional to the halo sizes from $z\sim0-3$ and the ratios slightly decrease toward $z\sim0$ and high stellar masses. Our studies are complementary to these studies. There are some new aspects in our work. We study the mass--angular momentum relation at high redshift. Moreover, while all previous studies have used abundance matching analysis, we use clustering analysis, which is independent of abundance matching analysis to estimate dark halo masses. We also compare our results with recent cosmological galaxy formation simulations.

The structure of our paper is as follows. In Section 2, we construct galaxy samples for this study. After measuring sizes in Section 3, we derive the stellar mass--disk size relation at each redshift bin in Section 4. The evolution of disk sizes is also discussed. In Section 5, we estimate dark halo masses from clustering analysis and abundance matching results. In Section 6, we present $j_{\star}$ and $m_{\star}$ estimates and compare them with recent cosmological galaxy formation simulations. Disk instabilities are also discussed. Conclusions are shown in Section \ref{seq_con}.

Throughout this paper, we adopt the cosmology ($\Omega_{\rm m}, \Omega_{\Lambda}, h$, $\sigma_8$) = (0.3, 0.7, 0.7, 0.8). Magnitudes are in the AB system \citep{oke83}. Galaxy sizes are given in the physical scale.

\section{Data and samples}\label{chap:data}
\subsection{Data}
We use data from the 3D-HST and CANDELS programs \citep{grogin11,koekeoer11, brammer12, skelton14}. \citet{skelton14} provide a photometric catalog of the 3D-HST and CANDELS imaging data for five sky fields (COSMOS, GOODS-North, GOODS-South, AEGIS, and UDS) with a total area of $\sim 900 \rm\ arcmin^2$. As these fields have wealthy available data of optical to near-infrared broadband photometry, one can obtain a precise spectral energy distribution (SED) for many high-redshift galaxies. The number of optical to near-infrared broadband filters ranges from 18 in UDS up to 44 in COSMOS. We make use of photometric redshift, stellar mass, and  star formation rates  (SFR), all of which are available through the 3D-HST Web site.\footnote[3]{http://3dhst.research.yale.edu}
Sources have been detected with $\tt{SExtractor}$ \citep{bertin96} from the combined F125W, F140W, and F160W images. Among the five fields we only use COSMOS, GOODS-South, and AEGIS fields because the clustering properties of galaxies in the remaining two fields appear to largely deviate from the cosmic average as detailed in Appendix.

Photometric redshifts have been determined from the $\tt{EAZY}$ \citep{brammer08} package, a public photometric redshift code. From the output catalog of $\tt{EAZY}$, we adopt $\tt{z\_peak}$ as photometric redshifts. Stellar masses and SFRs have been obtained by using the $\tt{FAST}$ code \citep{kriek09}. See \citet{skelton14} for details of the procedure. In this paper, we assume a \citet{chabrier03} initial mass function (IMF). From here, we take photometric redshifts as redshifts.

\subsection{Sample selection}
\begin{figure*}[tbp]
  \centering
      \includegraphics[width= 1.0\linewidth, trim=0 0 0 0, clip]{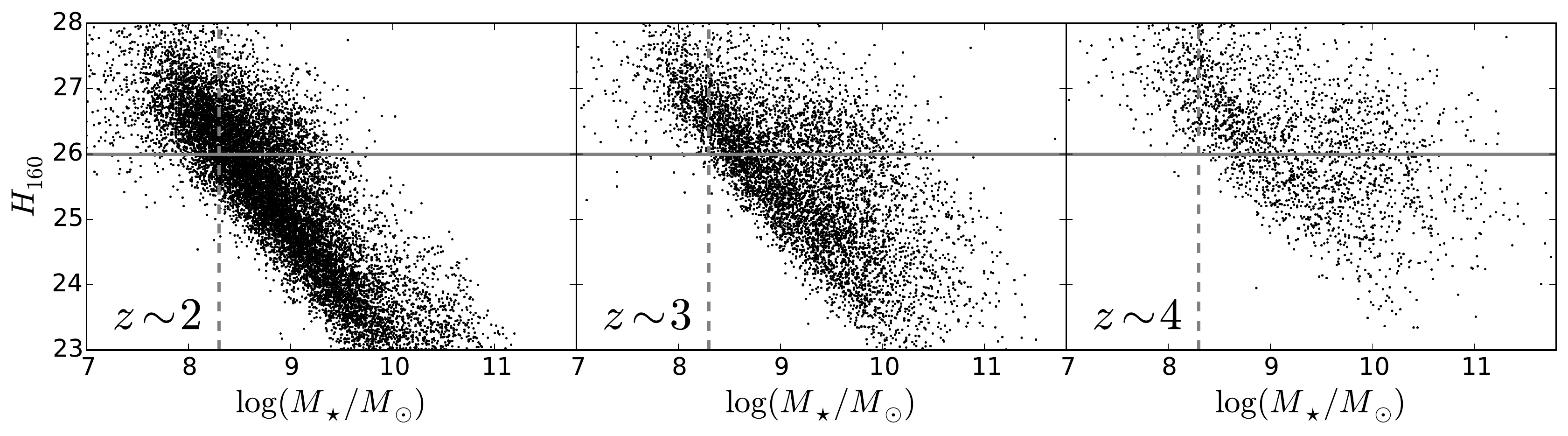}
  \caption{$H_{160}-M_{\star}$ diagram at $z\sim$ 2, 3, and 4 (left to right). The vertical dashed lines and horizontal solid lines indicate the stellar mass limits and the observed $H_{F160W}$ magnitude limits, respectively.}
  \label{fig2_1}
\end{figure*}

We limit our sample to $H_{160} < 26.0$, which is nearly equal to the 5$\sigma$ complete magnitude in the shallowest field COSMOS \citep{skelton14}. As size measurements need images with high signal to noise ratios (S/N), the 5$\sigma$ limit is marginally acceptable and slightly shallower compared to other size measurement studies \citep{vanderwel14, shibuya15}. 
Stellar masses are limited to $M_{\star} > 10^{8.3} M_{\odot}$. In the $H_{160}$--$M_{\star}$ diagram (Figure \ref{fig2_1}), stellar masses are largely complete down to $M_{\star} \simeq 10^{9.0} M_{\odot}$ for $z\sim2$ and down to $M_{\star} \simeq 10^{10}M_{\odot}$ for $z\sim3$ and 4. Below those values, our samples are biased toward low $M/L$ galaxies. We exclude galaxies with $M_{\star} > 10^{10.4} M_{\odot}$ from our samples for $z\sim3$ and $4$ because the number of galaxies is insufficient for clustering analysis.

\begin{figure*}[tbp]
  \centering
  	\includegraphics[width= 1.0\linewidth, trim=0 0 0 0, clip]{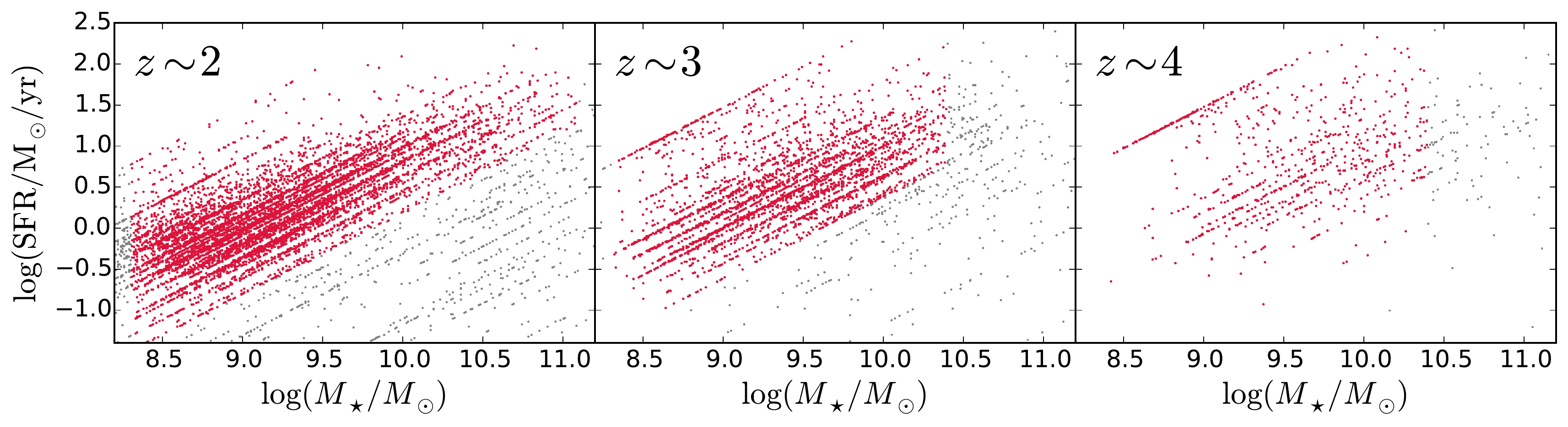}
  \caption{Star-formation rate vs. stellar mass diagram at $z\sim$ 2, 3, and 4 (left to right). Objects used in our study are shown in red.}
  \label{fig2_2}
\end{figure*}

We use the stellar mass--SFR diagram to remove quiescent galaxies. On the basis of the stellar masses and the SFRs obtained from the $\tt{FAST}$, we construct stellar mass--SFR diagrams for our samples, as shown in Figure \ref{fig2_2}. First, we fit the stellar mass--SFR distribution by a power law, which defines the main-sequence. At $z\sim$ 2 and 3, galaxies that lie above the $-2\sigma$ of the main-sequence are considered to be star-forming galaxies, where the standard deviation of the MS is $\sigma \simeq 0.33$ dex for both redshifts. For $z\sim4$, we remove galaxies that have small SFRs by eye. In this paper, we do not consider the effects of bulges because main sequence galaxies above $z\sim2$ have low B/T ratios \citep{brennan17}.

We exclude regions that have a shallow or deep exposure time for each field because clustering analysis requires images with a uniform depth. We also construct masks to avoid the vicinity of bright stars and diffraction spicks. For each redshift, we divide the entire sample into four ($z\sim$ 2) or three ($z\sim$ 3 and 4) subsamples according to stellar mass. The number of galaxies in the final samples is summarized in Table \ref{table2_1}.

\begin{deluxetable*}{ccrr}
\tabletypesize{\scriptsize}
\tablecaption{Number of star-forming galaxies for stellar mass subsamples\label{table2_1}} 
\tablewidth{0pt}
\tablehead{
  \colhead{$z$} & \colhead{$\log (M_{\star}/M_{\odot})$} & \colhead{ Number (Clustering) \tablenotemark{a}} & \colhead{ Number (Size ``success'') \tablenotemark{b}}}
\startdata
2.0 &$10.4 - 11.1$ & 264  & 198 \\
      & $9.7-10.4$ & 1086 & 870 \\
      & $9.0-9.7$ & 3267 & 2458\\
      & $8.3-9.0$ & 3173 & 1772\\
3.0 & $9.7-10.4$ & 805  & 560 \\
      & $9.0-9.7$ & 1596 & 1060 \\
      & $8.3-9.0$ & 838 & 412 \\
4.0 &$9.7- 10.4$& 273  & 161\\
      & $9.0-9.7$ & 348 & 176 \\
      & $8.3- 9.0$ & 133 & 70 
\enddata
\tablenotetext{a}{Number of star-forming galaxies used for clustering analysis.}
\tablenotetext{b}{Number of star-forming galaxies that have robust fitting parameters with  $\tt{GALFIT}$ detailed in Section \ref{seq_size_gal}}
\end{deluxetable*}

\section{Size measurements}\label{seq_size_measurements}
\subsection{Size measurements with $\tt{GALFIT}$} \label{seq_size_gal}
Galaxy sizes are measured for the F160W imaging data provided by the 3D-HST. 
Position, flux, half-light radius ($r_{\rm d}$), S\'ersic index ($n$), axis ratio ($q \equiv b/a$), and position angle are treated as free parameters to determine. 
In this paper, we use the half-light radius along the semi-major axis of the S\'ersic profile to define the size of galaxies.
We make $100$ pixels $\times$ 100 pixels cutout images around object galaxies before size measurement. We then run $\tt{GALFIT}$ \citep{peng02, peng10} on those cutout images, where neighbors are masked as not to perturb the fitting of the target galaxies. The masks are created from $\tt{SExtractor}$ segmentation maps.

As an initial guess of the free parameters, we use $\tt{SExtractor}$ output parameters given in the 3D-HST catalog. Results of $\tt{GALFIT}$ are not sensitive to initial values as long as they are not far from real values (H\"aussler et al., 2007). 
We vary individual parameters over the following ranges: $\Delta x,\ \Delta y < 3$ pixels, $0.3<r_{\rm d}<100$ pixel, $0.1<n<8$, $0.1<q<1$, where $\Delta x$ and $ \Delta y $ are the difference in the centroids between $\tt{SExtractor}$ and $\tt{GALFIT}$. We define galaxies whose best-fit parameters are within these ranges as ``success''. 
We only use ``success'' galaxies in the following analysis in Sections \ref{seq_size_measurements} and \ref{cap_mass_size}.
The number of ``success'' galaxies is summarized in  Table \ref{table2_1}. While we obtain robust structural parameters of only a part of our clustering sample, the average $\tt{SExtractor}$ sizes of the ``success'' sample and the entire sample are nearly equal. Thus we use the $\tt{GALFIT}$ sizes of the ``success'' sample as the representatives of the entire sample.

\subsection{Deriving $r_{\rm d}$ at rest-frame 5000$\AA$}
We derive $r_{\rm d}$ at the rest-frame $5000\AA$ at all redshifts. While we measure sizes in observed $1.6\mu m$ (F160W band), there exists a color gradient that depends on stellar mass and redshift. We obtain rest $5000\AA$ $r_{\rm d}$ by using the formula given in \citet{vanderwel14}:
\begin{eqnarray}
r_{\rm d} = r_{{\rm d}, F160W} \left(\frac{1+z}{1+z_p}\right)^{\Delta \log r_{\rm d}/\Delta \log \lambda}.
\end{eqnarray}
where $z_p$ is the ``pivot redshift"(2.2 for F160W) and the wavelength dependence is given by:
\begin{eqnarray}
\frac{\Delta \log r_{\rm d}}{\Delta \log \lambda} = -0.35 + 0.12z - 0.25 \log \left( \frac{M_{\star}}{10^{10} M_{\odot}}\right).
\end{eqnarray}
Although van del Wel et al. (2014) have only examined wavelength dependence over $0<z<2$, we extend this formula to $z \simeq 4$ because the redshift evolution of this relation looks linear as a function of redshift. In any case, the correction values at $z \sim$ 3 and 4 are relatively small.

\section{Stellar mass--size relation}\label{cap_mass_size}
\begin{figure*}[tbp]
  \centering
      \includegraphics[width= 1.0\linewidth, trim=0 0 0 0, clip]{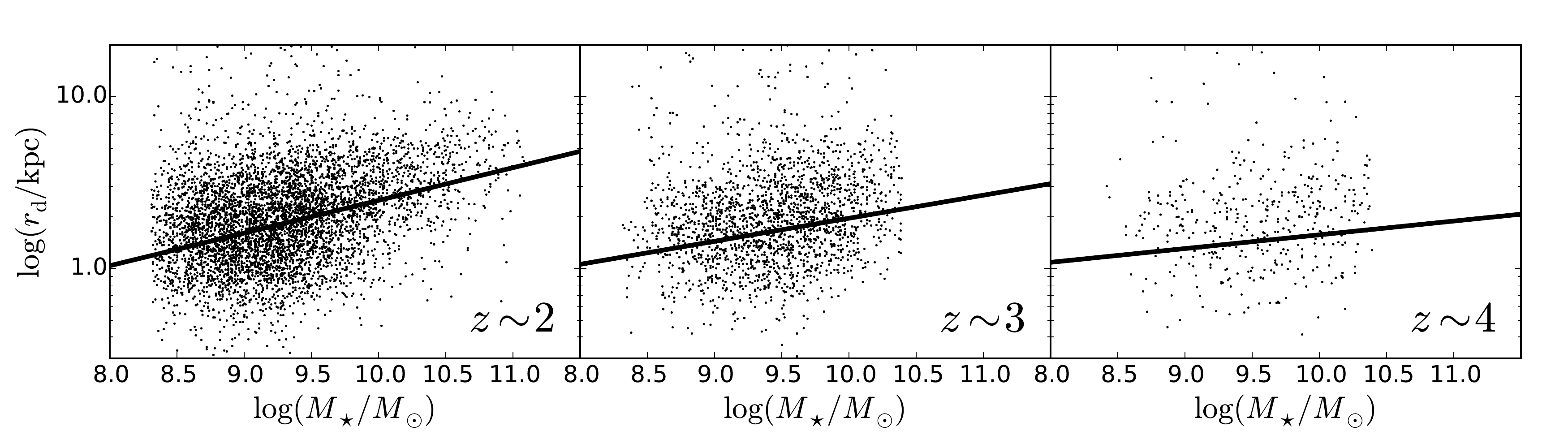}
  \caption{Stellar mass--size distribution of disk galaxies at $z\sim$ 2, 3, and 4 (left to right). The solid lines indicate the best-fit power-laws.}
  \label{fig4_1}
\end{figure*}

The stellar mass--size distributions of our star-forming galaxies are shown in Figure \ref{fig4_1}. 
In Section \ref{sec_modeling}, we analyze these distributions by modeling them with a power law. Then, we discuss the results in Section \ref{seq_sizeevolution}.

\subsection{Analytical Model of the stellar mass--size relation} \label{sec_modeling}
The stellar mass--size relation is usually modeled as a single power-law:
\begin{eqnarray}
\overline{r}_{\rm d} (M_{\star,10}) / {\rm kpc} =  A \cdot M_{\star,10}^{\alpha},
\end{eqnarray}
where $M_{\star,10} = M_{\star} / 1.0\times 10^{10} M_{\odot}$, and $\overline{r}_{\rm d}\ (M_{\star,10})$ is the median size at $M_{\star,10}$. For the size distribution at a given stellar mass, we adopt a log-normal distribution:
\begin{eqnarray}
p(r_{\rm d}|\sigma_{\ln r}, \overline{r}_{\rm d})dr_{\rm d} = \frac{1}{\sqrt{2\pi}\sigma_{\ln r} r_{\rm d}} \exp \left[ -\frac{(\ln r_{\rm d} - \ln \overline{r}_{\rm d})^2}{2\sigma^2_{\ln r}} \right] dr_{\rm d}, \nonumber \\
\end{eqnarray}
where $p(r_{\rm d}|\sigma_{\ln r}, \overline{r}_{\rm d})dr_{\rm d}$ is the probability density that a galaxy has a size between $(r_{\rm d}, r_{\rm d}+dr_{\rm d})$ at the given stellar mass, and $\sigma_{\ln r}$ is the dispersion of the distribution.
The reason for adopting a log-normal distribution comes from Equation~(\ref{eq1_1}). The disk size is proportional to the dimensionless spin parameter $\lambda$, and the distribution of $\lambda$ is well approximated by a log-normal distribution according to $N$-body simulations \citep{barnes87, bullock01}.

\begin{figure}[tbp]
  \centering
      \includegraphics[width= 0.9\linewidth, trim=0 0 0 0, clip]{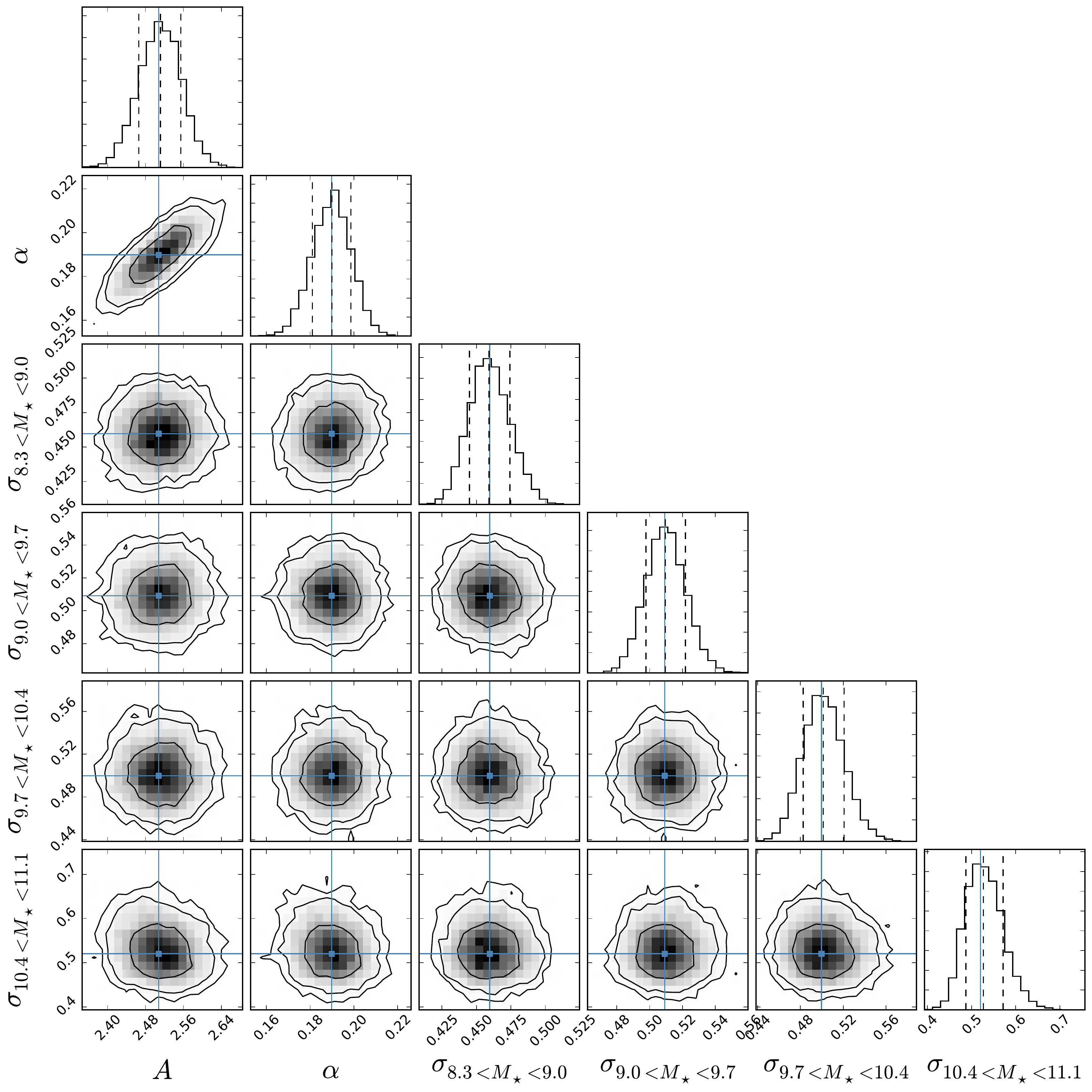}
      \includegraphics[width= 0.9\linewidth, trim=0 0 0 0, clip]{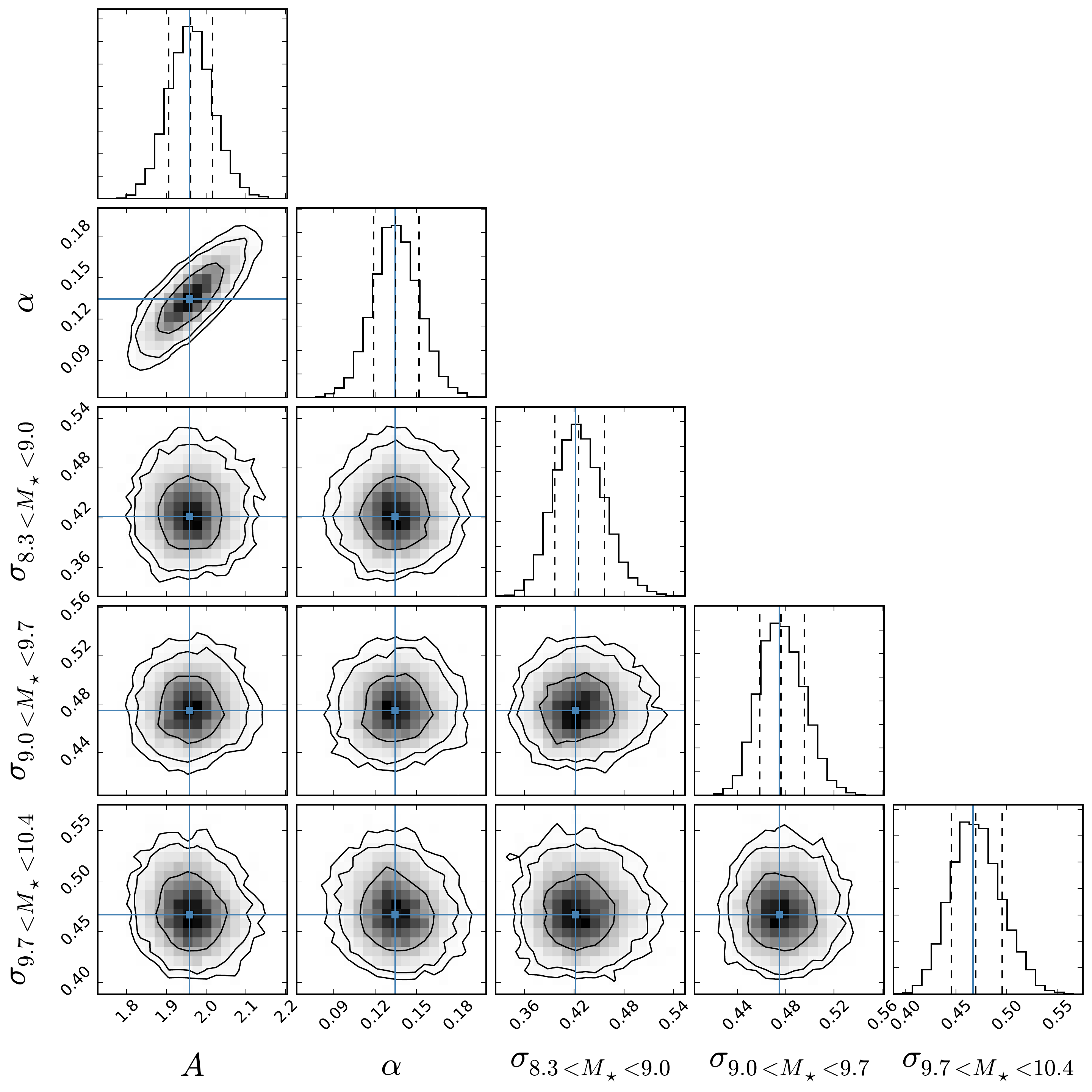}
      \includegraphics[width= 0.9\linewidth, trim=0 0 0 0, clip]{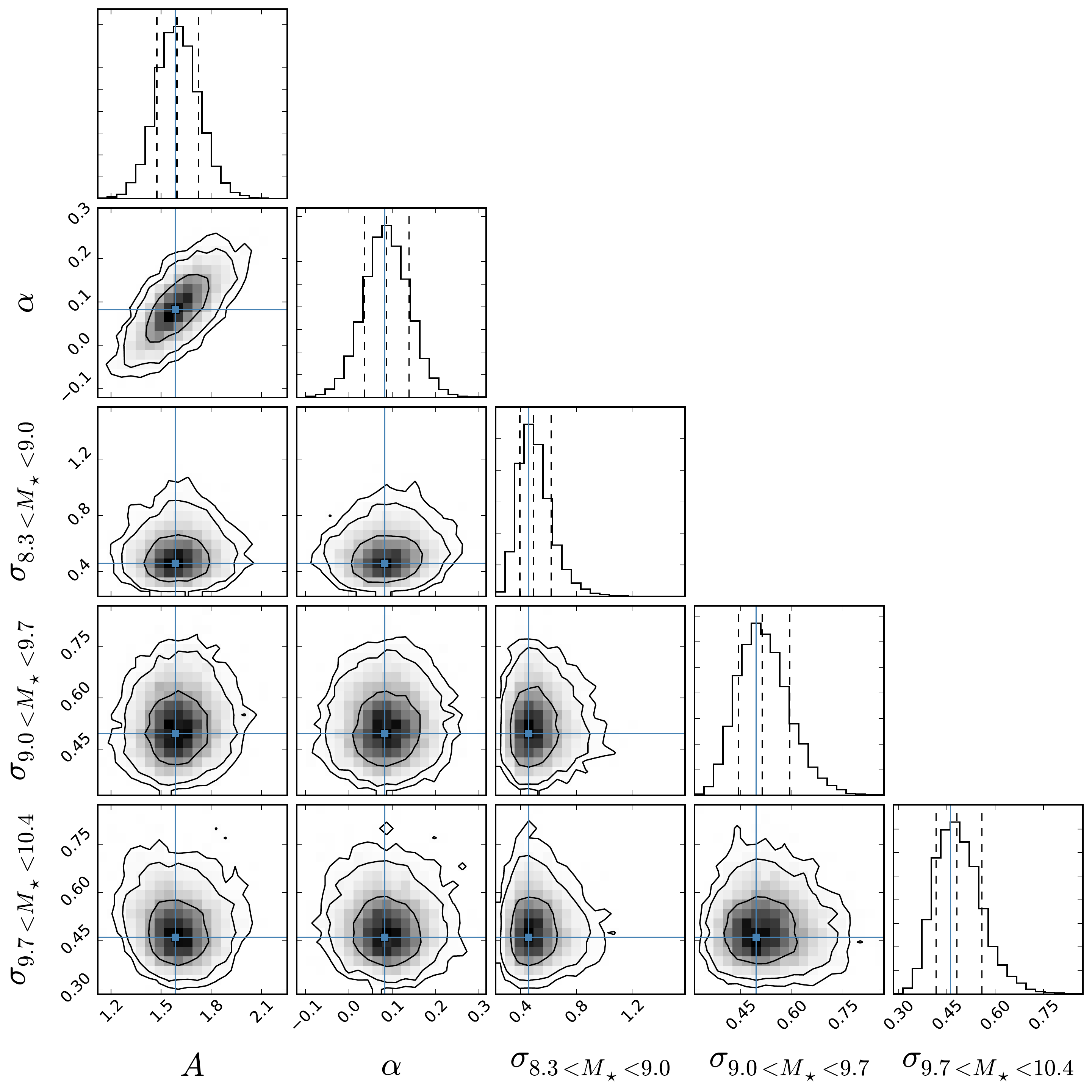}
  \caption{Sixty-eight percent, 96$\%$, and 99$\%$ confidence intervals for individual parameters at $z\sim2$, 3, and 4 (top to bottom). The top panel of each column shows the probability distribution function of each parameter. The solid blue lines indicate the median values.}
  \label{fig4_2}
\end{figure}

We assume that each of the observed disk sizes has a gaussian error:
\begin{eqnarray}
g ( x |\delta r_{\rm d})dx = \frac{1}{\sqrt{2\pi} \delta r_{\rm d}} \exp \left(- \frac{ x^2}{2 \delta r_{\rm d}^2} \right)dx,
\end{eqnarray}
where $g ( x | \delta r_{\rm d}) dx$ is the probability density that a galaxy has a intrinsic disk size between $x$ and $x+dx$.
The probability of observing $(r_{\rm d}, \delta r_{\rm d})$ assuming the log-normal distribution $p(r_{\rm d}|\sigma_{\ln r}, \overline{r}_{\rm d})$ is given by the convolution of the two functions:
\begin{eqnarray}
(p * g)(r_{\rm d}) = \int p(x) g(r_{\rm d}-x) dx.
\end{eqnarray}
We use the 1$\sigma$ error in $\tt{GALFIT}$ as $\delta r_{\rm d}$.
For each redshift, the free parameters of this model are given by $\textbf{P} = (A, \alpha, \sigma_{\ln r,i})$, where, $i$ denotes $i$-th subsample; we assume that different stellar mass bins have different $\sigma_{\ln r}$ values. We have six free parameters at $z\sim2$, and five free parameters at $z\sim$ 3 and 4.
We use the maximum likelihood estimation (MLE) to determine these parameters, where the estimated parameters make the observed $r_{\rm d}$ distribution the most probable. For subsample $i$ at a given redshift, the likelihood function is defined as
\begin{eqnarray}
\mathcal{L}_{i} = \prod_{j=1}^N (p * g)(r_{{\rm d},j}|\sigma_{\ln r,i}, \overline{r}_{\rm d}),
\end{eqnarray}
where $j$ represents the $j$-th object. We determine the parameter set $\textbf{P}$ that maximizes the likelihood function $\mathcal{L} \equiv  \prod \mathcal{L}_{i}$. The best-fit values are listed in Table \ref{table4_1}. We use the $\tt{scipy.optimize}$ package and the L-BFGS-B algorithm \citep{zhu97} to find the maximizing point. The uncertainties in the parameters are estimated by the Markov Chain Monte Carlo (MCMC) sampling. MCMC is a powerful algorithm to approximate multi-dimensional parameters using a Markov chain. We use the python package $\tt{emcee}$ \citep{foreman13} to run MCMC.
In Figure \ref{fig4_2}, we show for each parameter the best-fit values and the 68$\%$, 95$\%$, and 99$\%$ confidence intervals. This figure is made using the public python package $\tt{corner}$ \citep{foreman16}.

\begin{deluxetable*}{ccccccc}
\tabletypesize{\scriptsize}
\tablecaption{Best-fit parameters of the stellar mass--size relation\label{table4_1}} 
\tablewidth{0pt}
\tablehead{
  \colhead{$z$} & \colhead{$A$} & \colhead{ $\alpha$} & \colhead{$\sigma_{8.3<M_{\star}<9.0}$ }  & \colhead{$\sigma_{9.0<M_{\star}<9.7}$ } & \colhead{ $\sigma_{9.7<M_{\star}<10.4}$ } & \colhead{$\sigma_{10.4<M_{\star}<11.1}$}}
\startdata
2.0      & $2.51^{+0.03}_{-0.05}$ & $0.19^{+0.01}_{-0.01}$  & $0.46^{+0.01}_{-0.01}$ &  $0.51^{+0.01}_{-0.01}$   &   $0.50^{+0.02}_{-0.02}$ &   $0.53^{+0.03}_{-0.05}$ \\
3.0      &$1.94^{+0.06}_{-0.05}$  & $0.14^{+0.01}_{-0.03}$ & $0.42^{+0.03}_{-0.03}$ &  $0.47^{+0.02}_{-0.02}$   &   $0.47^{+0.02}_{-0.02}$ & \dots\\
4.0      & $1.57^{+0.11}_{-0.13}$  & $0.08^{+0.05}_{-0.05}$ & $0.45^{+0.18}_{-0.05}$ &  $ 0.51^{+0.08}_{-0.07}$   &   $0.47^{+0.09}_{-0.06}$ & \dots
\enddata
\end{deluxetable*}

\subsection[\bf Size evolution]{Size evolution} \label{seq_sizeevolution}
The evolution of $A$, $\alpha$, and $\sigma_{\ln r}$ are shown in Figure \ref{fig4_3}. In this Section, we discuss the evolution of each parameter in detail.

\subsubsection{Median size evolution}
The size evolution at a fixed stellar mass is generally parameterized as $(1+z)^{-\beta_z}$, where $\beta_z$ is a constant expressing the strength of evolution (evolution slope). The top panel of Figure \ref{fig4_3} represents the median size evolution of disk star-forming galaxies at $M_{\star} = 1.0 \times 10^{10} M_{\odot}$. The solid blue line shows the best-fit function over $z\sim2-4$: $\overline{r}_{\rm d} (M_{\star,10}) / {\rm kpc} =6.88 (1+z)^{-0.91 \pm 0.01}$. \citet{allen16} have measured the size evolution of a mass-complete sample $(\log (M_{\star}/M_{\odot}) > 10)$ of star-forming galaxies over redshifts $z = $ 1 $-$ 7, to find that the average size at a fixed mass of $\log (M_{\star}/M_{\odot}) = 10.1$ is expressed by $r_{\rm d} = 7.07(1+z)^{-0.89\pm 0.01}$. The slope we find is in agreement with \citet{allen16}'s value.  \citet{shibuya15} have also measured the stellar mass--median circularized size evolution of star-forming galaxies with $9.0 < \log (M_{\star}/M_{\odot}) < 11.0$ at $0<z<6$. The gray dotted line represents the average circularized half-light radius from their samples with the gray region showing the 16th and 84th percentiles. The evolution slope is consistent with our result. The difference in the amplitude is largely due to the different definition of galaxy sizes. We also note that \citet{shibuya15} have used the Salpeter IMF \citep{salpeter55} to derive stellar masses.

However, $\beta_z = 0.91 \pm 0.01$ is slightly steeper than the value by \citet{vanderwel14}. They have studied a mass complete sample of star-forming galaxies and have found $(1+z)^{-0.75}$ at a $\log (M_{\star}/M_{\odot}) = 10.7$ over the redshift range $0 < z < 3$. As their method of size measurements is the same as ours, we attribute this discrepancy to the difference in the redshift range. The evolution slope of star-forming galaxies appears to become steeper above $z\sim$ 2 or 3. \citet{allen16}'s sample also shows steeper slopes at higher redshifts \citep[See Figure 3 of][]{allen16}. 

As size evolution is closely related to the evolution of hosting dark matter halos, $\beta_z$ contains information of dark matter halos. From Equation~(\ref{eq1_1}), when $r_{\rm d}/r_{\rm 200}$ is constant irrespective of $z$ and $M_{\rm dh}$, $r_{\rm d}$ is given by
\begin{eqnarray}
r_{\rm d} &\propto& H(z)^{-1}V_{\rm c} \label{eq4_2_1} \\
&\propto&  H(z)^{-2/3}M^{1/3}_{\rm dh}, \label{eq4_2_3}
\end{eqnarray}
where $V_{c}$ is the circular velocity of dark matter halos.
The Hubble parameter as a function of $z$, $H(z) = H_0 \sqrt{\Omega_{\rm m} (1+z)^3 + \Omega_{\Lambda}}$, is approximated as $H(z) \propto (1+z)^{1.5}$. According to Equations~(\ref{eq4_2_1}) and (\ref{eq4_2_3}), $r_{\rm d} \propto (1+z)^{-1.5}$ means evolution at a constant circular velocity and $r_{\rm d} \propto (1+z)^{-1.0}$ means evolution at a constant virial mass \citep{ferguson04}. 
The $\beta_{z} = 0.91$ is close to the prediction for a constant virial mass.  

\subsubsection{Slope evolution} \label{sub_slope}
\begin{figure}[tbp]
  \centering
      \includegraphics[angle=0,width=0.7\linewidth, trim=0 0 0 0, clip]{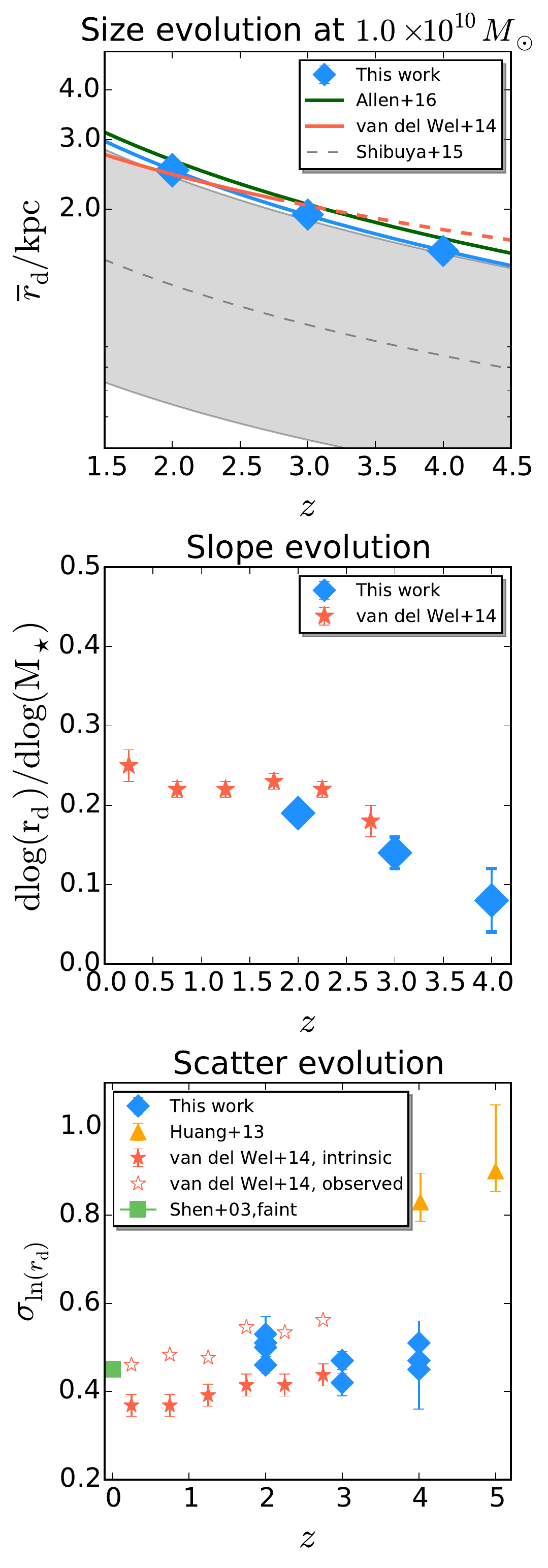}
  \caption{Redshift evolution of the stellar mass--size relation of star-forming galaxies. Top: the size evolution at $M_{\star} = 1.0 \times 10^{10} M_{\odot}$. The blue diamond symbols indicate the results obtained in this paper, and the solid blue line shows the best-fit power law. The green solid line shows the average size of star forming galaxies from \citet{allen16} at $10^{10.1}M_{\odot}$. The red solid line indicates the size evolution of late-type galaxies from \citet{vanderwel14} at $10^{9.75} M_{\odot}$, and the red dashed line is its extrapolation. The gray dotted line and the shaded region indicate the median circularized size and the 16th and 84th percentiles distribution of star-forming galaxies with $9.5<\log M_{\star}/M_{\odot}<10.0$ \citep{shibuya15}. Middle: slope evolution. The blue and red symbols represent our galaxies and late-type galaxies from \citet{vanderwel14}, respectively. Bottom: the intrinsic scatter evolution from this work and previous studies. The blue symbols represent our galaxies. The orange symbols represent LBGs from \citet{huang13} at $z\sim$ 4 and 5. The filled and open red symbols show the late-type galaxies of \citet{vanderwel14}. The green symbol shows the SDSS galaxies of \citet{shen03} at the faint end.}
  \label{fig4_3}
\end{figure}

The middle panel of Figure \ref{fig4_3} shows the slope evolution in the stellar mass--size relation ($\alpha$). The slope evolution of the stellar mass--size relation for late-type galaxies was first investigated by \citet{vanderwel14}. They have found that the slope has nearly a constant value $\simeq 0.2$ over the redshift range $0<z<3$. Similarly \citet{allen16} have found  $\alpha = 0.15\pm 0.01$ for star-forming galaxies at $1<z<2.5$. Our results are consistent with those of \citet{vanderwel14} and \citet{allen16} at $z\sim$ 2 and 3, however, being slightly lower at $z\sim 4$.

The slope evolution of the stellar mass--size relation is determined as a combination of the slope of the stellar mass--halo mass relation and the slope of the disk size--halo size relation. In this paper, We have measured all three slopes. We will discuss the relation between the three slopes in Section \ref{seq_ang}.

\subsubsection{Scatter evolution}
We present the evolution of the intrinsic scatter in the bottom panel of Figure \ref{fig4_3}: here,  ``intrinsic'' means that measurement errors have been removed. The scatter for local galaxies is generally small. \citet{shen03} have found $\sigma_{\ln r_{\rm d}} \sim 0.3$ for both late-type and early type galaxies from SDSS. This result has also been ascertained by the result of \citet{courteau07}, $\sigma_{\ln r_{\rm d}} \sim 0.3$, for local spiral galaxies. These studies have been extended by \citet{vanderwel14} to the high-redshift universe and they have found that the intrinsic scatter dose not strongly evolve since $z\sim 2.75$ for both late-type and early-type galaxies. In their study, the scatter for late type galaxies is $0.16-0.19$ $\rm dex$, which is comparable to the result of \citet{shen03} and  \citet{courteau07}. We extend \citet{vanderwel14}'s study up to $z\sim4$, and find that the intrinsic scatter is constant with $0.4-0.6$ over $z\sim2-4$.

The scatter of $\lambda$ has been specifically investigated by $N$-body simulations and found to be $\sigma_{\lambda} \sim 0.5$ \citep{bullock01}. Thus the disk formation model of Equation~(\ref{eq1_1}) naively predicts that the intrinsic scatter of sizes is $\sim 0.5$.

The results for local galaxies imply that the size scatter is smaller than that of the spin parameter $\lambda$. To explain the observed small scatters, some mechanisms are needed. One possible mechanism is bulge growth. The growth of bulges increases the specific angular momentum of disks and thus expands disk sizes. Low-spin galaxies selectively grow their bulges. Some kind of disk instability and feedback has also been proposed which remove galaxies with low-spin and high-spin halos.

Our result, $\sigma_{\ln r_{\rm d}} \sim 0.4$ $-$ 0.6, is comparable with the scatter of the log-normal distribution of $\lambda$. This implies that for star-forming galaxies at $z\sim$ 2 $-$ 4 the size scatter at a given stellar mass is fully explained by the scatter of $\lambda$. Our result, however, does not agree with the large scatters, $\sigma_{\ln r_{\rm d}}\sim0.8-0.9$, found by \citet{huang13} for the size--UV luminosity relations of $z\sim4-5$ LBGs. This may suggest that the UV luminosity--halo mass relation of LBGs has a considerably large scatter.

\section{Halo mass estimates}\label{seq_halo}
In this Section we estimate the masses of the dark matter halos hosting our galaxies by using two independent methods: clustering analysis and abundance matching technique. Clustering analysis utilizes the large scale clustering amplitude of observed galaxies to obtain their hosting dark matter halo masses. Clustering analysis is a popular way to estimate hosting dark matter halo masses, however the mass estimates in this paper have relatively large errors because the sizes of individual subsamples are not so large. To test the results of the clustering analysis, we use abundance matching technique, which connects the stellar mass of galaxies to that of dark matter halos. While abundance matching can easily estimate hosting dark matter halo masses, it does not consider that different galaxy types have different stellar mass dark halo mass ratios. We briefly explain the two methods and show the obtained dark matter halo masses.

\subsection{Clstering analysis}
\subsubsection{Angular correlation function}

We compute the two point angular correlation functions (ACFs), $\omega_{\rm true}(\theta)$, of star-forming galaxies. Here, we assume all of our galaxies as central galaxies. 
The observed ACFs, $\omega_{\rm obs}(\theta)$, are measured by counting the number of unique pairs of observed galaxies and comparing it with what is expected from random samples.
We adopt the estimator proposed by \citet{landy93}:
\begin{eqnarray}
\omega_{\rm obs}(\theta) = \frac{ DD(\theta) - 2DR(\theta) + RR(\theta)}{RR(\theta)},
\end{eqnarray}
where $DD(\theta)$, $DR(\theta)$, and $RR(\theta)$ are the normalized numbers of galaxy-galaxy, galaxy-random and random-random pairs, respectively, with separation $\theta$. We generate 1000 times as many random points as the number of galaxies accounting for the geometry of the observed area and the masks. The formal error in $\omega_{\rm true}$ is given by
\begin{eqnarray}
\sigma_{\omega} = \sqrt{[1+\omega_{\rm obs}]/DD(\theta)}.
\end{eqnarray}
We assume a power low parameterization for the ACF,
\begin{eqnarray}
\omega_{\rm true} (\theta) = A_{\omega} \theta^{-\beta}. \label{eq1}
\end{eqnarray}
We fix $\beta = 0.8$ following previous studies \citep[e.g.][]{peebles75, ouchi01, ouchi04, ouchi10, foucaud03, foucaud10, harikane16}. 

\begin{figure*}[tbp]
  \centering
      \includegraphics[width= 1\linewidth, trim=0 0 0 0, clip]{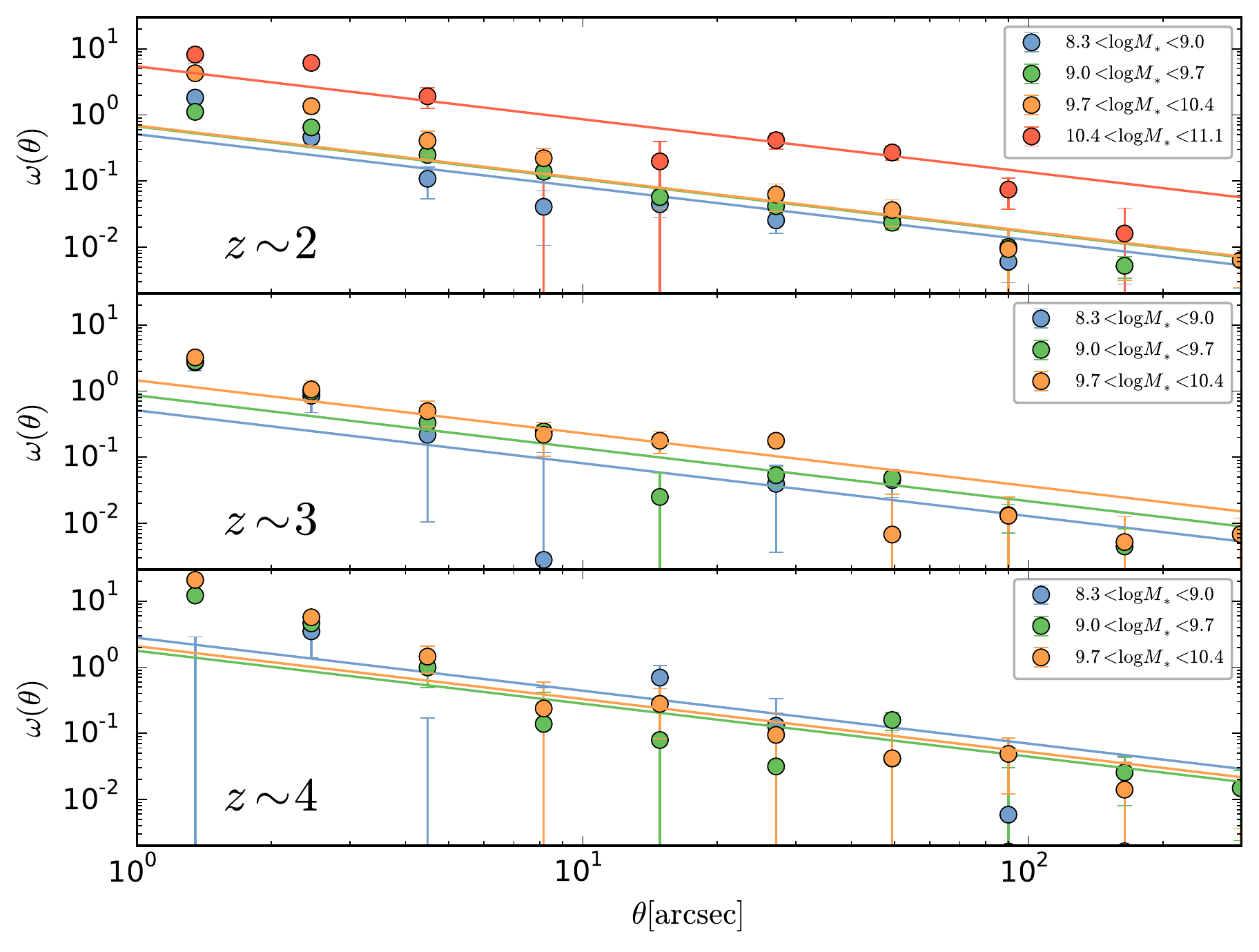}
  \caption{Angular correlation functions of star-forming galaxies at $z\sim$ 2, 3, and 4 (top to bottom). Data points and the best-fit power laws are color-coded by the stellar mass range.}
  \label{fig5_1}
\end{figure*}

It is known that $\omega_{\rm obs}$ is underestimated because we only use a finite survey area. This is compensated by introducing an integral constraint (IC) \citep{groth77}:
\begin{eqnarray}
\omega_{\rm true} = \omega_{\rm obs} + \rm{IC}.
\end{eqnarray}
The IC value depends on the size and shape of the survey area, and is estimated using a random catalog: 
\begin{eqnarray}
{\rm IC} = \frac{\sum_i RR(\theta_i) \omega_{\rm true}(\theta_i)}{\sum_i RR(\theta_i)} =  \frac{\sum_i RR(\theta_i) A_{\omega} \theta^{-\beta}}{\sum_i RR(\theta_i)}.
\end{eqnarray} 
Because the three 3D-HST fields used in this paper have almost the same size, we obtain nearly the same IC value (${\rm IC}_{\rm GOODS-S} = 0.016A_{\rm \omega}$, ${\rm IC}_{\rm COSMOS} = 0.013A_{\rm \omega}$, and ${\rm IC}_{\rm AEGIS} = 0.010A_{\rm \omega}$).
The amplitude $A_{\omega}$ is estimated through the ACFs of the three fields by minimizing $\chi^2$:
\begin{eqnarray}
\chi^2 = \displaystyle \sum_{i,\,j={\rm fields}} \frac{[A_{\omega} \theta_{i}^{-\beta} - (\omega_{{\rm obs},j}(\theta_{i}) + {\rm IC}_j)]^2}{\sigma_{\omega,j}^2(\theta_{i})},
\end{eqnarray}
where ${\rm IC}_j$, $\omega_{{\rm obs}, j}$, and $\sigma_{\omega,j}^2(\theta)$ denote the IC, observed ACF, and errors in field $j$, respectively. We use data at $\theta > 10"$ for fitting because at $\theta < 10"$ the contribution of the one halo term cannot be ignored. 
In figure \ref{fig5_1} we plot the ACFs of our subsamples with the best-fit power laws.

Then we estimate the spatial correlation function, $\xi(r)$, from the measured ACFs and the redshift distribution of galaxies.  The spatial correlation function is usually assumed to be a single power low as
\begin{eqnarray}
\xi(r) = \left( \frac{r}{r_0} \right)^{-\gamma},
\end{eqnarray}
where $r_0$ is the correlation length and $\gamma$ is the slope of the power low. These parameters are related to those of the two point angular correlation function via the Limber transform \citep{peebles80, efstathiou91}.
\begin{eqnarray}
\beta &=& \gamma - 1, \\
A_{\omega} &=& \frac{r_0^{\gamma} B[1/2, (\gamma - 1)] \int^{\infty}_0 dz N(z)^2 F(z) D_{\theta}(z)^{1-\gamma} g(z)}{[\int^{\infty}_0 N(z)dz]^2 }, \nonumber  \\ \\
g(z) &=& \frac{H_0}{c}(1+z)^2 \{1 + \Omega_{\rm m} z + \Omega_{\Lambda}[(1+z)^{-2} -1]\}^{1/2},
\end{eqnarray}
where $D_{\theta}(z)$ is the angular diameter distance, $N(z)$ is the redshift distribution of galaxies, $B$ is the beta function, and $F(z)$ describes the redshift evolution of $\xi(r)$. $F(z)$ is often modeled as $F(z) = [(1+z)/(1+z_{\rm c})]^{-(3+\overline{\epsilon})}$ with $\epsilon = -1.2$ \citep{roche99}, where $z_{\rm c}$ is the characteristic redshift of galaxies. We assume that the clustering evolution is fixed in comoving coordinates over the redshift range in question.

\subsubsection{Galaxy biases and halo masses}
To understand the relation between galaxies and hosting dark matter halos we use the halo model of \citet{sheth01}, which is obtained from the ellipsoidal collapse model. 
In the model of \citet{sheth01} the bias factor of dark halos, $b_{\rm dh}$, is given by
\begin{eqnarray}
b_{\rm dh} = 1 + \frac{1}{\delta_{\rm c}} \left[ \nu'^{2} + b \nu'^{2(1-c)} - \frac{\nu'^{2c} / \sqrt{a}}{\nu'^{2c} + b(1-c)(1-c/2)} \right], \nonumber  \\
\end{eqnarray}
where $\nu' = \sqrt{a} \nu$, $a = 0.707$, $b = 0.5$, $c = 0.6$, and $\delta_{\rm c} = 1.69$ is the critical amplitude above which overdense regions collapse to form a virialized object. Here, $\nu$ is defined as
\begin{eqnarray}
\nu = \frac{\delta_{\rm c} }{\sigma(M,z)} =  \frac{\delta_{\rm c} }{D(z) \sigma(M,0)},
\end{eqnarray}
where $D(z)$ is the linear growth factor, $\sigma(z)$ is the mass rms. of the smoothed density field.  We calculate $D(z)$ by the formula of \citet{carroll92} and $\sigma(M,0)$ using an initial power spectrum of a power law index $n=1$ and the transfer function of \citet{bardeen86}.
Then we define the linear galaxy bias, which is the relation between the clustering amplitude of galaxies and that of dark matter halos, at a large scale ($=8 h^{-1}_{100} \rm\ Mpc$) as
\begin{eqnarray}
b_{\rm g} = \sqrt{\frac{\xi_{\rm g}( r = 8 h^{-1}_{100} \rm\ Mpc)}{\xi_{\rm DM} (r=8 h^{-1}_{100} \rm\ Mpc)}} = \sqrt{\frac{[8 h^{-1}_{100} \rm\ Mpc / r_0]^{-\gamma}}{\xi_{\rm DM} (r=8 h^{-1}_{100} \rm\ Mpc)}}, \nonumber \\
\end{eqnarray}
where $\xi_{\rm DM} (r=8 h^{-1}_{100} \rm\ Mpc)$ is the dark matter spatial correlation function. We calculate  $\xi_{\rm DM} (r=8 h^{-1}_{100} \rm\ Mpc)$ using the non-linear model of \citet{smith03}.
Assuming that the galaxy bias at large scales is almost the same as the halo bias ($b_{\rm g} \simeq b_{\rm dh}$), we obtain an estimate of dark halo masses. The correlation length and the estimated halo masses are summerized in Table \ref{table5_1}.  

\begin{deluxetable*}{ccrcccc}
\tabletypesize{\scriptsize}
\tablecaption{Summary of the clustering analysis and the abundance matching analysis\label{table5_1}} 
\tablewidth{0pt}
\tablehead{
  \colhead{$z$} & \colhead{$\log (M_{\star}/M_{\odot})$} & \colhead{$N$} & \colhead{$A_{\omega}\rm [arcsec^{0.8}]$} &  \colhead{$r_0[h^{-1} \rm Mpc]$}  & \colhead{$\log(M_{\rm dh,CL}[h^{-1} M_{\odot}])$} & \colhead{$\log(M_{\rm dh,AM}[h^{-1} M_{\odot}])$}}
\startdata
2.0 &10.58& 264  & $5.40^{+0.96}_{-0.96}$ &    $12.30^{+1.18}_{-1.25}$   &   $13.37^{+0.10}_{-0.12}$ & 12.23\\
      & 9.94 & 1086 & $0.69^{+0.25}_{-0.25}$ &  $ 3.92^{+0.73}_{-0.87}$   &   $11.69^{+0.32}_{-0.56}$ & 11.79\\
      & 9.30 & 3267 & $0.67^{+0.07}_{-0.07}$ &  $ 3.86^{+0.21}_{-0.23}$   &   $11.66^{+0.11}_{-0.13}$ & 11.51\\
      & 8.72 & 3173 & $0.51^{+0.08}_{-0.08}$ &  $ 3.31^{+0.28}_{-0.30}$   &   $11.32^{+0.18}_{-0.23}$ & 11.30\\
3.0 & 9.93 & 805  & $1.45^{+0.31}_{-0.31}$ &   $ 5.18^{+0.58}_{-0.65}$   &   $11.92^{+0.17}_{-0.23}$ & 11.81\\
      & 9.37 & 1596 & $0.86^{+0.15}_{-0.15}$ &  $ 3.87^{+0.36}_{-0.39}$   &   $11.40^{+0.17}_{-0.21}$ & 11.53\\
      & 8.78 & 838 & $0.51^{+0.31}_{-0.31}$ &  $ 2.90^{+0.87}_{-1.18}$   &   $10.79^{+0.56}_{-1.60}$ & 11.29 \\
4.0 &10.01& 273  & $2.08^{+0.93}_{-0.93}$ &   $ 5.57^{+1.27}_{-1.56}$   &   $11.79^{+0.31}_{-0.56}$ & 11.78\\
      & 9.37 & 348 & $1.77^{+0.72}_{-0.72}$ &  $ 5.09^{+1.06}_{-1.28}$   &   $11.64^{+0.30}_{-0.51}$ & 11.45 \\
      & 8.82 & 133 & $2.78^{+1.74}_{-1.74}$ &  $ 6.54^{+2.03}_{-2.75}$   &   $12.03^{+0.38}_{-0.91}$ & 11.22
\enddata
\end{deluxetable*}

\subsection{Abundance Matching}
In order to reinforce the results of the clustering analysis, we also use abundance matching analysis, which connects the number density of galaxies to that of dark halos to estimate the hosting dark halo mass for a given stellar mass. We adopt the abundance matching result of \citet{behroozi13}. Many researchers that study the angular momentum retention factor adopt the abundance matching analysis of \citet{dutton10} and \citet{behroozi13} to estimate halo masses \citep[e.g.][]{fall12, burkert16}. This makes easy to compare our results with previous results of angular momentum studies. The estimated halo masses are also summerized in Table \ref{table5_1}.  

\begin{figure}[tbp]
  \centering
      \includegraphics[width= 1.0\linewidth, trim=0 0 0 0, clip]{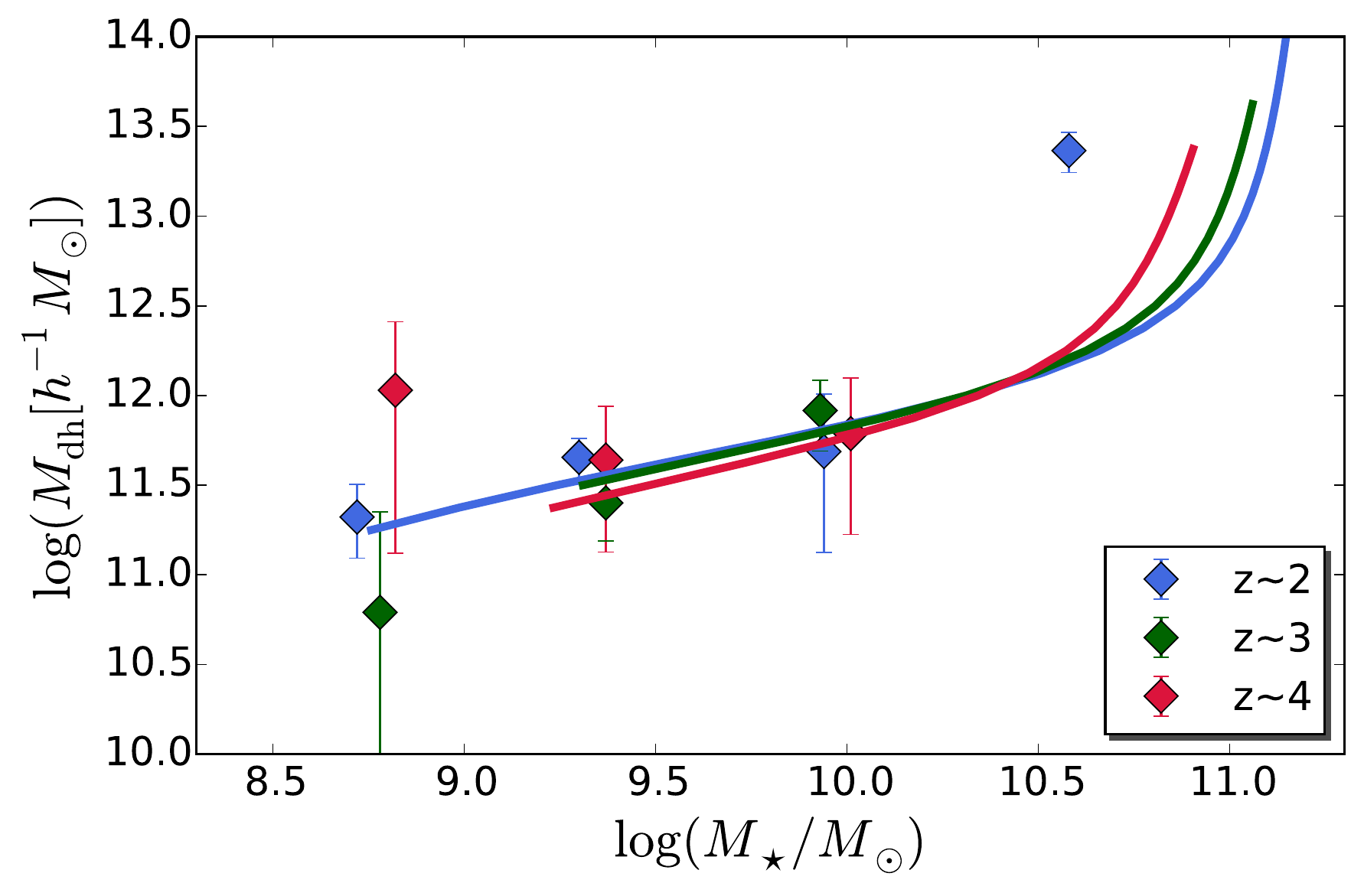}
  \caption{Dark matter halo mass as a function of stellar mass obtained from clustering analysis and abundance matching technique at $z\sim$ 2, 3, and 4. The diamonds indicate the results of clustering analysis, while the solid lines indicate the results of the abundance matching of \citet{behroozi13}.}
  \label{fig5_2}
\end{figure}

Figure \ref{fig5_2} shows a comparison of the estimated dark matter halo masses. The estimated dark matter halo masses by the two independent methods are consistent within the error bars except for the highest stellar mass bins at $z\sim 2$. This makes the results of the clustering analysis more plausible. In the following Section, we display the results based on the both methods.

\section{Angular momentum}\label{seq_ang}
\subsection{Estimation of the specific angular momentum} \label{sec_estimate}
In this Section, we briefly explain the way to estimate the disk specific angular momentum. As already mentioned in Section \ref{sec1}, the disk size of a galaxy reflects its specific angular momentum. According to the model of \citet{mo98}, the specific angular momentum of disk galaxies with an exponential profile ($n=1$) is given by:
\begin{eqnarray}
j_{\rm d} =   \frac{\sqrt{2}}{1.678} r_{\rm d} m_{\rm d} \lambda^{-1} r_{200}^{-1} f_{\rm c}(c_{\rm vir})^{1/2} f_{\rm R}(\lambda, c_{\rm vir}, m_{\rm d},j_{\rm d})^{-1}.  \nonumber \\ \label{eq6_1}
\end{eqnarray}
If we assume $r_{\rm d}$ as the half-light radius of a S\'ersic index $n$, we can expand this equation to:
\begin{eqnarray}
j_{\rm d} &=&   f_n(n)^{-1} r_{\rm d} m_{\rm d} \lambda^{-1} r_{200}^{-1} f_{\rm c}(c_{\rm vir})^{1/2} f_{\rm R}(\lambda, c_{\rm vir}, m_{\rm d},j_{\rm d})^{-1}, \nonumber \\ \label{eq6_2} \\
f_n(n) &=& \frac{\sqrt{2} \Gamma(2n) \kappa^n}{\Gamma(3n)},
\end{eqnarray}
where $\Gamma$ is a gamma function, and $\kappa$ is well approximated by
\begin{eqnarray}
\kappa &=& 2n - \frac{1}{3} + \frac{4}{405n} + \frac{46}{25525n^2} \nonumber  \\
&\quad&+ \frac{131}{1148175n^3} +  \mathcal{O}(n^{-4})\ (n>0.36),\\
\kappa &=& 0.01945 - 0.8902 n + 10.95 n^2 - 19.67n^3 \nonumber \\
&\quad&+ 13.43 n^4\ (n<0.36)
\end{eqnarray}
\citep{ciotti99,macarthur03}. The full functional forms of $f_{\rm c}$ and $f_{\rm R}$ are given in \citet{mo98}. The values of $\lambda$ and $c_{\rm vir}$ are well determined by $N$-body simulations \citep{vitvitska02, davis09, prada12, puebla16}. We adopt $(\lambda, c_{\rm vir}) = (0.035,4.0)$ throughout the examined redshift range ($z \sim 2-4$). From the dark matter halo masses estimated in Section \ref{seq_halo}, we can calculate $m_{\rm d}$ and $r_{200}$, where $r_{200}$ is calculated by
\begin{eqnarray}
r_{200} = \left(\frac{GM_{\rm dh}}{100 H(z)^2}\right)^{1/3}.
\end{eqnarray}
Combined with $n$ and $r_{\rm d}$ measured in Sections \ref{seq_size_measurements} and \ref{cap_mass_size}, we can estimate $j_{\rm d}$.

\subsection[\bf Mass--angular momentum relation]{Mass--angular momentum relation}
\subsubsection[\bf Average $j_{\rm d}/m_{\rm d}$ ratio and its evolution]{Average $j_{\rm d}/m_{\rm d}$ ratio and its evolution}
Figure \ref{fig6_3} shows the angular momentum retention factor of star-forming galaxies as a function of hosting halo mass. We find $j_{\star}/m_{\star} = 0.77\pm0.06$ from clustering analysis and $j_{\star}/m_{\star} = 0.83\pm0.13$ from abundance matching at $z \sim$ 2, 3, and 4. No strong redshift evolution is confirmed. As we mention in Section \ref{sec1}, $j_{\star}/m_{\star} = 1$ means that the angular momentum is fully conserved and $j_{\star}/m_{\star} < 1$ means that galaxies lose their specific angular momentum during their formation and evolution.

\citet{fall12} have investigated kinematical structure for about 100 bright early and late-type galaxies at $z\sim0$. They have found that late-type galaxies typically have $j_{\rm d}/m_{\rm d} \simeq 0.6$ and early-type galaxies have $j_{\rm d}/m_{\rm d} \simeq 0.1$. A small $j_{\rm d}/m_{\rm d}$ value has also been reported by \citet{dutton12}. They have calculated angular momentum retention factor as a function of halo mass by constructing the mass models \citep{dutton11} tuned to observed scaling relations for SDSS galaxies. They have obtained a constant value $j_{\rm d}/m_{\rm d} = 0.61^{+0.13}_{-0.11}$ with halo masses $10^{11.3}M_{\odot} \lesssim M_{\rm dh} \lesssim 10^{12.7} M_{\odot}$. Our values at $z\sim$ 2, 3, and 4 are in rough agreement with these local values for late-type galaxies within errors.

There exist a few studies that have investigated the mass--angular momentum relation at high redshifts. Recently, \citet{burkert16} have investigated the relation for $\sim360$ star-forming galaxies at $z\sim0.8-2.6$, among which about 100 are at $z\sim2$, by H$\alpha$ kinematics based on KMOS and SINS/zC-SINF surveys. They have found $j_{\rm d}/m_{\rm d} = 1.0$ with a statistical uncertainty of $\pm0.1$ and a systematic uncertainty of $\pm0.5$. This $j_d/m_d$ value is consistent with our result at $z\sim2$.

We then compare our results with those of \citet{huang17} and \citet{somerville17}. These authors have derived disk size to halo size ratios $(r_{\rm d}/r_{\rm dh})$ as a function of stellar mass over $z \sim 0$ and 3 using the CANDELS data and mapping stellar masses to halo masses with abundance matching. At $z\sim2$, the $r_{\rm d}/r_{\rm dh}$ ratios obtained by \citet{huang17} are consistent with ours, with values of $\sim0.03$ in the stellar mass range $10^9 M_{\odot} < M_{\star} < 10^{10.5} M_{\odot}$. We note that our method is very similar to theirs. Their definitions of disk sizes and halo sizes are the same as ours. They have used four abundance matching results 
including that of \citet{behroozi13} which we also use. On the other hand, \citet{somerville17} have obtained somewhat higher ratios of $r_{\rm d}/r_{\rm dh} \simeq 0.4$. They have adopted a different halo definition and also taken a different method to link stellar masses to halo masses; they have carried out ``forward modeling" where halos are taken from an $N$-body simulation and are assigned to stellar masses taking account of a random scatter. These differences may be a cause of the inconsistency in $r_{\rm d}/r_{\rm dh}$ estimates.
 
 To connect our study to those for low redshifts, we use Extended Press-Schechter (EPS) formalism \citep{bond91, bower91, lacey93}. The EPS formalism is able to calculate the conditional probability mass function ($f(M_{2}|M_{1})$) of $z=z_{2}$ descendant halos for a given halo mass ($M_{1}$) at a high-redshift ($z_{1}$) universe by following their merger histories. We set $M_{1} = 5.0 \times 10^{11}h^{-1}M_{\odot}$ and $z_{1} = 3.0$ to follow the evolution of our halos. The lower 68 and upper percentiles of $f(M_{2}|M_{1})$ at $z_{2}=0$ are $2.0\times 10^{12}h^{-1}M_{\odot}$ and $5.6 \times 10^{12}h^{-1}M_{\odot}$, respectively. This implies that some fraction of our galaxies are the progenitors of objects in the \citet{dutton12} sample in terms of mass growth.
 From the results we obtain, we can depict a unified view of the angular momentum evolution. 
Disk galaxies maintain high $j_{\rm d}/m_{\rm d}$ values during their evolution from cosmic noon to the present day, unless they lose angular momenta by some mechanisms like mergers and turn into early-type galaxies \citep{fall12}.

\subsubsection[\bf Halo mass dependence of $j_{\rm d}/m_{\rm d}$ and the slope of the size--stellar mass relation]{Halo mass dependence of $j_{\rm d}/m_{\rm d}$ and the slope of the size--stellar mass relation}

\begin{figure*}[tbp]
  \centering
      \includegraphics[angle=0,width= 1.0\linewidth, trim=0 0 0 0, clip]{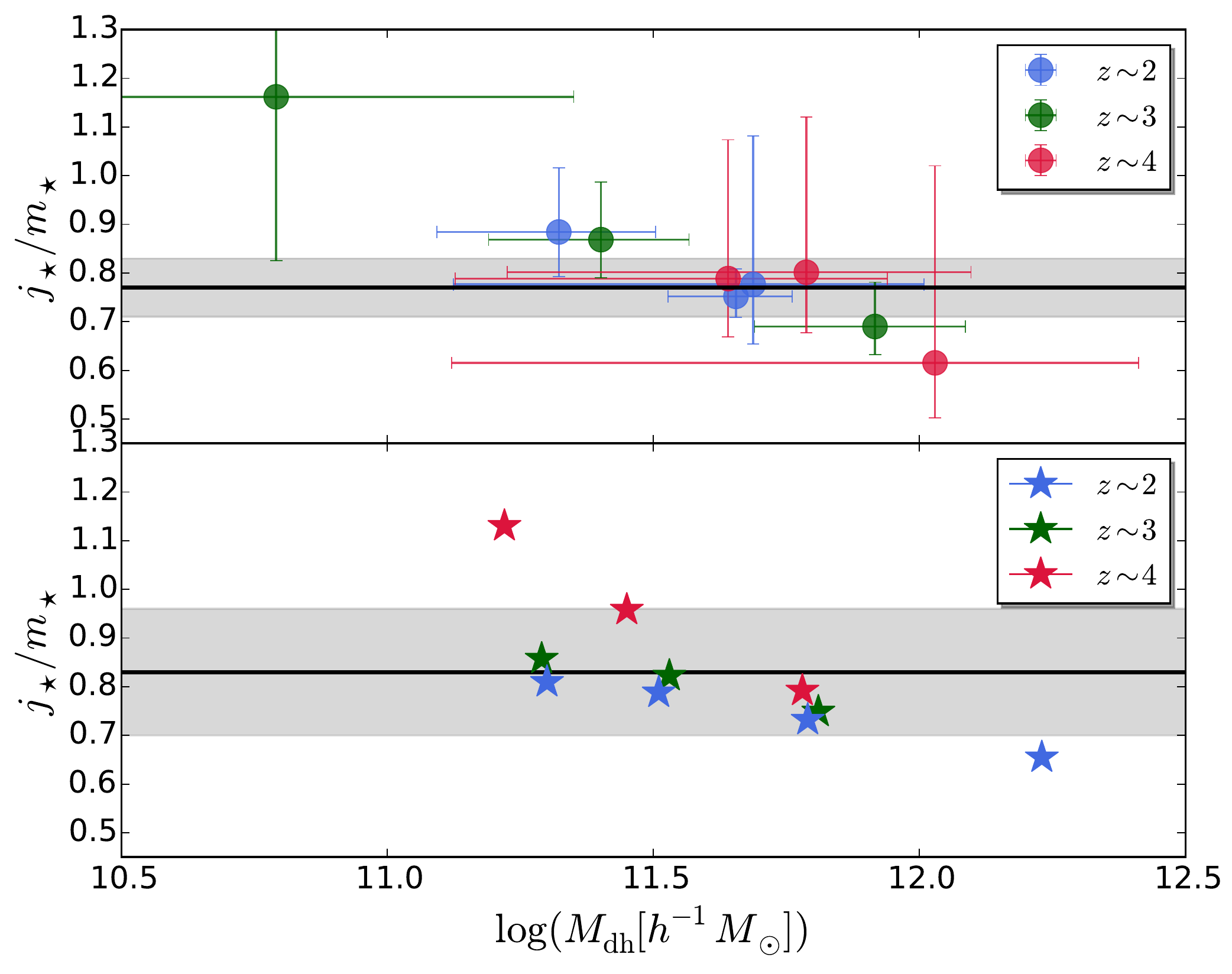}
  \caption{Angular momentum retention factor $j_{\star}/m_{\star}$ vs. $M_{\rm dh}$ for $z\sim$ 2, 3, and 4. The colored symbols in the top panel and the bottom panel indicate the results of clustering analysis and the results of abundance matching analysis, respectively. For each panel, the black solid line and the gray shaded region indicate the average of all estimates and its $1\sigma$ error, respectively.}
  \label{fig6_3}
\end{figure*}

When we introduce the disk size--halo mass relation in Equation~(\ref{eq4_2_3}), we assume that $r_{\rm d}/r_{\rm 200}$ is constant, which means that $j_{\star}/m_{\star}$ is constant irrespective of $z$ and $M_{\rm dh}$. However,  it appears from Figure \ref{fig6_3} that $j_{\star}/m_{\star}$ weakly depends on both $M_{\rm dh}$ and $z$. Similar dependencies have also been shown in \citet{huang17} and \citet{somerville17}: $r_{\rm d}/r_{\rm dh}$  weakly depends on both $M_{\rm dh}$ and $z$. We approximate the observed $j_{\star}/m_{\star}$--$M_{\rm dh}$ relation at each redshift by a power law, $j_{\star}/m_{\star} \propto M_{\rm dh}^{\gamma_{z}}$. We find $\gamma_{z} = -0.09\pm0.02$ for $z\sim2$, $\gamma_{z} = -0.13\pm0.01$ for $z\sim3$, and $\gamma_{z} = -0.29\pm0.02$ for $z\sim4$. A negative slope of $\gamma_{z} =  -0.19 \pm 0.04$ has also been obtained by \citet{burkert16} for $z\sim 0.8-2.6$ galaxies. With a non-zero slope $\gamma_{z}$, Equation~(\ref{eq4_2_3}) is replaced by:
\begin{eqnarray}
r_{\rm d} \propto H(z)^{-2/3} M_{\rm dh}^{\gamma_{z}+1/3}.
\end{eqnarray}
We also approximate the stellar mass--halo mass relation by a single power-law, $M_{\star} \propto M_{\rm dh}^{\epsilon}$: $\epsilon \simeq1.6$, from the abundance matching results of \citet{behroozi13}. By combining these two relations, we obtain the disk size--stellar mass relation:
\begin{eqnarray}
r_{\rm d} &\propto& M_{\star}^{1/3\epsilon + \gamma_{z}/\epsilon},\\
r_{\rm d} &\propto& M_{\star}^{0.2+0.6 \gamma_{z}}. \label{eq6_3}
\end{eqnarray}
The slope of the size--stellar mass relation of our galaxies is $\alpha$ = $0.19^{+0.01}_{-0.01}$ for $z\sim2$, $0.14^{+0.01}_{-0.03}$ for $z\sim3$, and $0.08^{+0.05}_{-0.05}$ for $z\sim4$ (see Section \ref{sub_slope}).
The result that $\alpha$ is less than 0.2 for all three redshifts is explained by the negative $\gamma_{z}$ values obtained above. We also find that the decrease in $\alpha$ from $z\sim3$ to $z\sim4$ is due to the decrease in $\gamma_{z}$.

Using a theoretical modified cooling model which includes disc instability, \citet{dutton12} have predicted a slightly negative $\gamma_{z}$ for high redshift disk galaxies, in qualitative agreement with our results. Their negative slope reflects the fact that the mass loading factor decreases with increasing of halo mass. While this model is not consistent with their empirical model at $z\sim0$, this model may be applicable to high redshifts. The possible decrease in $\gamma_{z}$ from $z\sim3$ to $z\sim4$ found above may imply that feedback processes also change in this redshift range.

 As already seen in Figure \ref{fig4_3}, \citet{vanderwel14} have reported constant disk size--stellar mass slopes ($\sim 0.2$) since $z\sim2-0$. From the model of Equation~(\ref{eq6_3}), this implies that the angular momentum--halo mass relations are also flat. This is quite in agreement with the empirical results of \citet{dutton12} at the present-day universe. Thus Equation~(\ref{eq6_3}) well represents the relation between angular momentum and disk size.

\subsection{Comparison with galaxy formation models} \label{sec_conparison}
As the kinematics of galaxies provides us with important constraints on galaxy formation and evolution as well as do other global properties like stellar mass, star-formation rate, and metallicity, many modelers have attempted to reproduce the kinematic structures of galaxies. Early attempts concerning angular momentum with hydrodynamical simulations were in trouble with reproducing observations. They suffered from unexpected angular momentum loss. In those simulations, most of the angular momentum of galaxies was transferred to the background hosting halos. As a result, compact disk galaxies were produced \citep[e.g.][]{navarro91,navarro94}. This problem is known as the ``angular momentum catastrophe''. 

This problem has been considerably improved by high-resolution hydrodynamical simulations with a proper treatment of feedback processes \citep{robertson06,governato07,scannapieco08}. 
In recent years, many galaxy formation simulations have succeeded in reproducing the mass--angular momentum relation for both early-type and late-type galaxies in the present-day universe \citep{genel15, teklu15}. On the other hand, at high redshifts, there do not exist theoretical studies that compare with observational data. It is still unknown that these simulations are able to reproduce the observed mass--angular momentum relation beyond $z\sim 1$. Here, we first compare our observational angular momentum results with those of some galaxy formation simulations \citep{sales12, pedrosa15, stevens16}. 

\begin{figure*}[tbp]
  \centering
      \includegraphics[width= 1\linewidth, trim=0 0 0 0, clip]{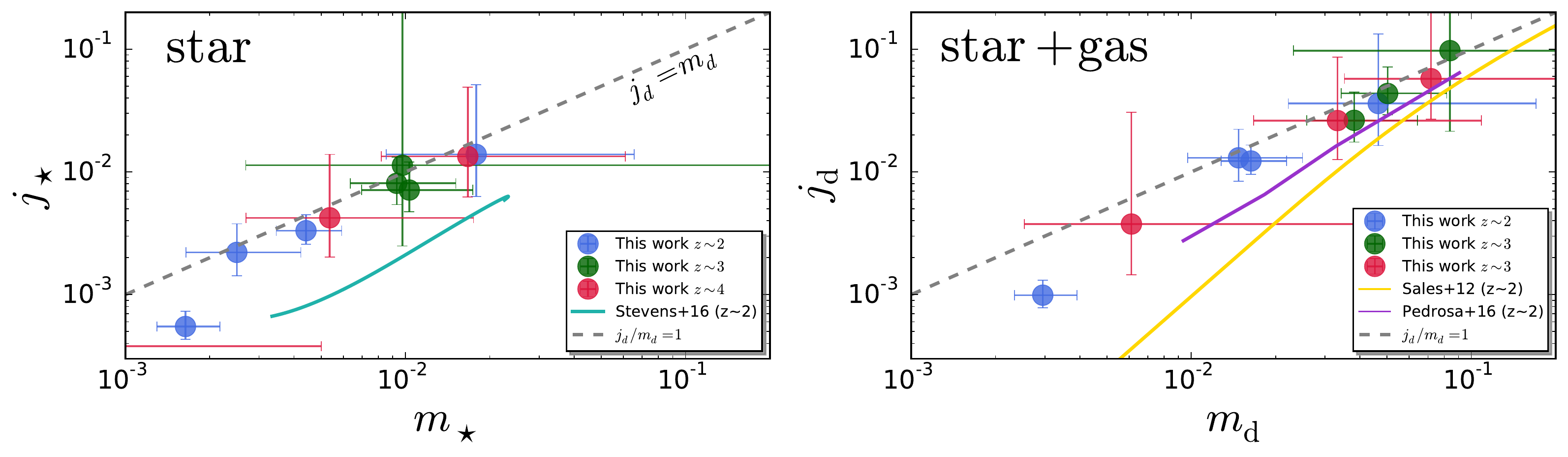}
      \includegraphics[width= 1\linewidth, trim=0 0 0 0, clip]{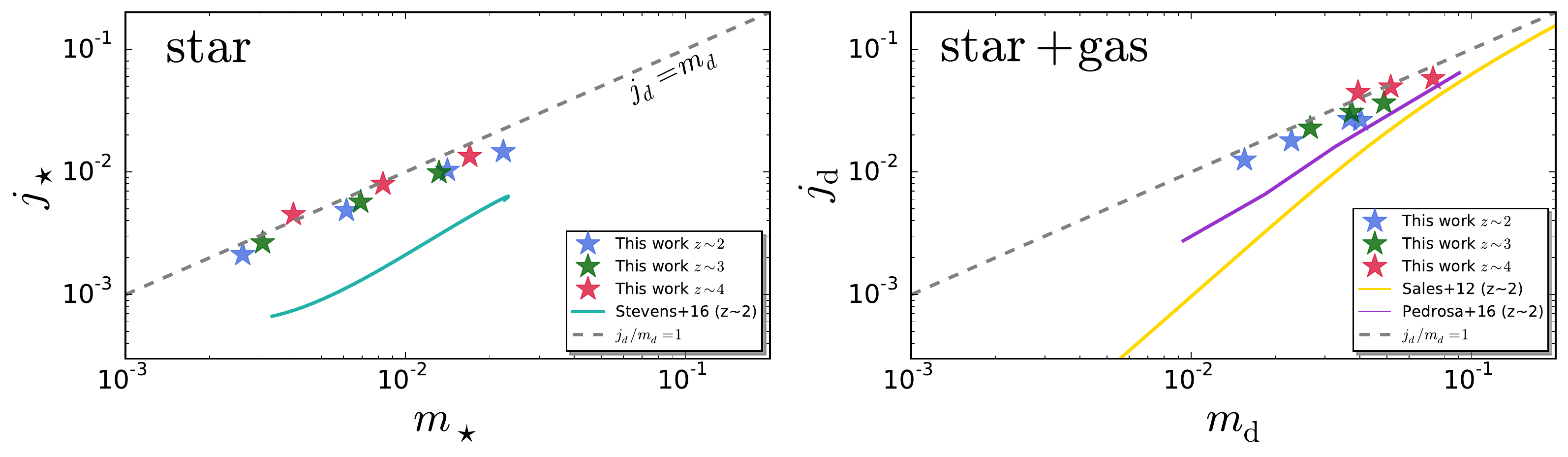}
  \caption{Observed mass--angular momentum relation compared with three hydrodynamical and semi-analytic galaxy formation simulations. The left panels show the relation for the stellar component and the right panels for the stellar plus gas component. The colored symbols in the top panels are the results obtained from clustering analysis and those in the bottom panels are from abundance matching analysis. The solid cyan lines on the left panels indicate the semi-analytical galaxy formation simulation of \citet{stevens16} at $z\sim2$. The solid purple and yellow lines on the right panels indicate the hydrodynamical galaxy formation simulations of \citet{sales12} and \citet{pedrosa15}, respectively, at $z\sim 2$. The gray dashed lines indicate the line of angular momentum conservation.}
  \label{fig6_4}
\end{figure*}

In Figure \ref{fig6_4}, we compare the mass--angular momentum distribution of star-forming galaxies  obtained from clustering analysis and abundance matching analysis with predictions from hydrodynamical and semi-analytical galaxy formation models at $z\sim2$. To directly compare with two models which give only stellar plus gas properties, we also estimate the entire disk masses by correcting for gas masses using the gas fraction estimates given in \citet{schinnerer16}. They have investigated the gas masses for 45 massive star-forming galaxies observed with ALMA at redshifts of $z\sim3-4$. We extend their results to lower mass and lower redshift by the prediction of 2-SFM (2 star formation mode) model \citep{sargent14}. We correct $m_{\star}$ and $j_{\star}$ by the same factor assuming that the stellar and gas disks have the same $j$ value. The right panel of Figure \ref{fig6_4} shows the baryonic disk mass--angular momentum relation. 

\citet{sales12} have presented baryonic mass--angular momentum relations with various types of feedback from large cosmological $N$-body/gasdynamical simulations at $z\sim 2$. They have found that regardless of the strength of the feedback process $m_{\rm d}$ vs. $j_{\rm d}$ follows the same relation (the yellow solid lines in Figure \ref{fig6_4}). When strong feedbacks push out most of the baryons from the galaxies, both $m_{\rm d}$ and $j_{\rm d}$ are reduced. \citet{pedrosa15} have also analyzed the mass--angular momentum relation by decomposing disks and bulges with cosmological hydrodynamical simulations at $z\sim0-2$. They have found no significant evolution since $z\sim2 $ to $z\sim 0$. The relation for total baryonic components at $z\sim2$ is shown in Figure \ref{fig6_4}.

\citet{stevens16} have presented a semi-analytical model D{\scriptsize ARK} S{\scriptsize AGE}, which is designed for specific understanding of angular momentum evolution. They have investigated the evolution of the stellar mass--specific angular momentum relation over $0<z<4.8$. The solid cyan lines in Figure \ref{fig6_4} indicate the predicted mass--angular momentum relation at $z\sim2$. Here, we assume the abundance matching results by \citet{behroozi13} to map stellar mass to dark halo mass and an analytical model by \citet{fall12}, which connects dark matter halo mass to their halo angular momentum:
\begin{eqnarray}
j_{\rm vir} = 4.23 \times 10^{4} \lambda \left(\frac{M_{\rm vir}}{10^{12} M_{\odot}} \right)^{2/3} {\rm km\,s^{-1}\,kpc}.
\end{eqnarray}
Note that as the \citet{fall12}'s model uses cosmological parameters at the present day $(c_{\rm vir}=9.7,\Delta_{\rm vir}=319)$, we replace them with values $(c_{\rm vir}=4.0,\Delta_{\rm vir}=200)$.

All of these simulations predict specific angular momenta systematically smaller than our values from both dark matter halo mass estimation methods. Our relations are almost parallel to the line of angular momentum conservation (dotted gray lines in Figure \ref{fig6_4}) regardless of mass scales, however, the simulations predict smaller specific angular momenta and the deviations are large for smaller $m_{\star}$ and $m_{\rm d}$. While the star+gas plots appear to have smaller deviations than those of the star only plots, note that we ignore a possible difference in the distribution of gases and stars within galaxies. In other words, we assume that gases and stars have the same specific angular momentum. However, \citet{brook11_1} have shown that the angular momentum distributions of stars and $\rm H_{I}$ gases are different, with $\rm H_{I}$ gases having a tail of high angular momentum. Indeed, extended $\rm H_{I}$ gas disks are found in intermediate \citep{puech10} and high redshift \citep{daddi10} galaxies. Gases beyond star-forming regions serve as a high angular momentum reservoir \citep{brook11_1}. These gases should have a larger specific angular momentum than stars. In this case, the gaps on the right panels in Figure \ref{fig6_4} become larger.

These deviations imply that these simulations produce too small disk sizes at high redshifts. Some mechanisms that increase disk specific angular momentum at high redshifts may be needed. For example, \citet{brook12} have proposed that selective ejection of low angular momentum material from galaxies leads to a redistribution of angular momentum. This explains the difference in the distribution of angular momentum between dark matter halos and visible galaxies: dark matter halos have a large low angular momentum tail, while observed galaxies do not. This process reproduces large bulge-less high angular momentum galaxies.
 
 Whether or not these feedback related mechanisms are enough to solve the deviations seen in Figure \ref{fig6_4} is still unknown. More detailed observations and simulations are needed.

\subsection{Disk instability}
The angular momentum of disks is also closely related to their global instabilities. Disks can be unstable against bar mode instability, because low angular momentum material forms a bar  \citep{shen03}. \citet{efstathiou82} have investigated this kind of instabilities for a exponential disk embedded in a variety of halos using $N$-body simulations and found a stellar disk is globally unstable against bar formation under the criterion:
\begin{eqnarray}
\epsilon_{\rm m} \equiv \frac{V_{\rm max}}{(GM_{\rm d} / r_{\rm d})^{1/2}} \lesssim 1.1, \label{eq6_9}
\end{eqnarray}
where $V_{\rm max}$ is the maximum rotation velocity of the disk. The threshold for gaseous disks is $\epsilon_{\rm m} \simeq 0.9$. According to \citet{mo98}, for a NFW halo, this criterion is well approximated by
\begin{eqnarray}
\lambda' < m_{\rm d}, \label{eq6_10}
\end{eqnarray}
where $\lambda' \equiv \lambda j_{\rm d} / m_{\rm d}$. 

We note that the criteria of Equations~(\ref{eq6_9}) and (\ref{eq6_10}) are not strict. \citet{guo11} have proposed an alternative criterion, $V_{\rm max} < \sqrt{GM_{\rm d}/3r_{\rm d}}$, which reflects that $V_{\rm max}$ of the real dark matter halo systems is smaller than that of ideal systems. In this paper, we use Equation~(\ref{eq6_10}).

\begin{figure}[tbp]
  \centering
      \includegraphics[width= 1\linewidth, trim=0 0 0 0, clip]{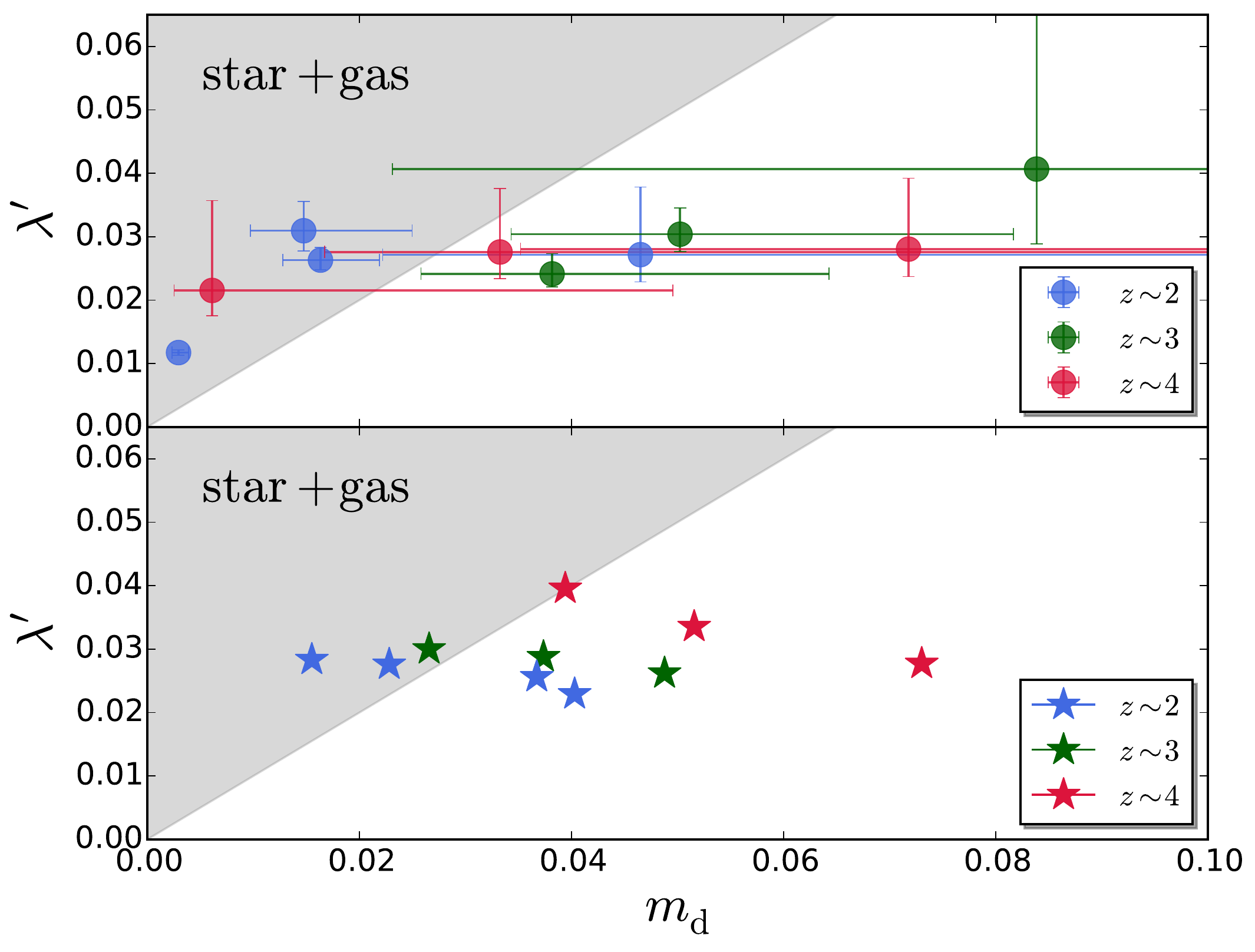}
  \caption{Diagram of $\lambda'$ vs. $m_{\rm d}$ at $z\sim$ 2, 3, and 4. The colored symbols in the top panel and the bottom panel indicate results from clustering analysis and abundance matching analysis, respectively. Galaxies in the gray shaded regions are unstable against bar-mode instability.}
  \label{fig6_5}
\end{figure}

We show in Figure \ref{fig6_5} the distribution in the $\lambda'$--$m_{\rm d}$ plane of our star-forming galaxies over $z \sim$ 2 $-$ 4. We find most of the data points to be near the line of instability over the entire redshift range regardless of the method to estimate dark halo masses. This implies some fractions of $z \sim 2-4$ galaxies may be dynamically changing the disk structure toward forming a bar and a bulge through bar formation.

To compare with local spiral galaxies, we assume $\lambda = 0.04$ and $j_{\rm d}/m_{\rm d} \simeq 0.6$ \citep{fall12} in the present-day Universe. Then, the average value of $\lambda'$ is estimated as 0.024. The abundance matching result of \citet{behroozi13} predicts $m_{\rm d}$ lower than 0.024 in a wide range of halo mass.
This displays that local spiral galaxies appear to be more stable than high redshift galaxies.

We have to keep in mind again that we should take into account a possible difference in angular momentum between gases and stars mentioned in Section \ref{sec_conparison}. In this case, the plots in Figure \ref{fig6_5} will move to more stable regions.

Other than the global instability, there exist scenarios that form bars and bulges \citep{mo10}. For example, an interaction with a massive perturber leads to a bar-like structure \citep{noguchi87}. In addition to this, the migration of giant clumps, which are created by local Toomre Q instabilities \citep{toomre64}, grows a bulge. Global instability may be one of the ways to explain galaxies with bars or bulges in the local Universe.

\section{Conclusion}\label{seq_con}
In this paper, we have used the 3D-HST GOODS-South, COSMOS, and AEGIS imaging data and galaxy catalog to analyze the relation between the ratio of the disk stellar mass to the halo mass, $m_\star \equiv M_\star/M_{\rm dh}$, and the fraction of the dark halo angular momentum transferred to the stellar disk, $j_\star \equiv J_\star/J_{\rm dh}$ for 11738 star-forming galaxies over the stellar mass range $8.3 < \log(M_{\star}/M_{\odot})< 11.1$ at $z\sim$ 2, 3, and 4.
For each redshift, we have divided the catalog into several $M_{\star}$ bins and infer $M_{\rm dh}$ by two independent methods, clustering analysis and abundance matching, to obtain an average $m_{\star}$ value for each bin. We have confirmed that the two mass estimators give consistent results. For our objects we have also measured effective radii $r_{\rm d}$ at rest 5000\AA\ with $\tt{GALFIT}$, and combined them with $m_{\star}$ and $M_{\rm dh}$ estimates to obtain $j_{\star}$ by applying  \citet{mo98} analytic model of disk formation. The followings are the main results of this paper.
\\\\
(i) We have found the median size evolution of disk star-forming galaxies $\overline{r}_{\rm d} (M_{\star,10}) / {\rm kpc} =6.88 (1+z)^{-0.91 \pm 0.01}$ at $M_{\star} = 1.0\times 10^{10}M_{\odot}$. This redshift evolution is in agreement with the results by \citet{allen16} and \citet{shibuya15}.
We have also analyzed the slope of the disk size--stellar mass relation. While the slope is consistent with the results by \citet{vanderwel14} at $z\sim2$, we have found that the slope becomes shallower beyond $z\sim$ 2. The scatter of $r_{\rm d}$--$M_{\star}$ relation is $\sigma_{\ln r_{\rm d}} \sim 0.4$ $-$ 0.6 over the redshift range examined, which is comparable with the scatter of the log-normal distribution of $\lambda$. 
\\\\
(ii) We have obtained the angular momentum retention factor $j_{\star}/m_{\star}$ averaged over mass and redshift to be $\simeq 0.77\pm0.06$ from clustering analysis and  $\simeq 0.83\pm0.13$ from abundance matching. These values are in rough agreement with those of local late-type galaxies by \citet{fall12} and those of star-forming galaxies at $z\sim0.8-2.6$ by \citet{burkert16}.
\\\\
(iii) Contrary to the star-forming galaxies at the present-day universe, $j_{\star}/m_{\star}$ appears to decrease with halo mass especially when abundance matching is used as the mass estimator. Combined with the slope of the $M_{\star}$--$M_{\rm dh}$ relation, this negative slope of the $j_{\star}/m_{\star}$--$M_{\rm dh}$ relation explains the shallow ($<0.2$) slopes of the $r_{\rm d}$--$M_{\star}$ relation obtained in this paper. We have also found a possible decrease in the $j_{\star}/m_{\star}$--$M_{\rm dh}$ slope from $z\sim2$ to $z\sim4$, which may imply that feedback processes also change over this redshift range.
\\\\
(iv) We have for the first time compared the observed mass--angular momentum relation with those of the recent galaxy formation simulations at $z\sim2$ by \citet{sales12}, \citet{pedrosa15}, and \citet{stevens16}. We have found that all of these simulations predict specific angular momenta systematically smaller than our values, which implies that these simulations produce too small disks at high redshifts while reproducing local measurements. We have also found that a significant fraction of our galaxies appear to be unstable against bar formation.

\section*{Acknowledgments}
This work is based on observations taken by the 3D-HST Treasury Program (GO 12177 and 12328) with the NASA/ESA HST, which is operated by the Association of Universities for Research in Astronomy, Inc., under NASA contract NAS5-26555.
This work is supported by KAKENHI (16K05286) Grant-in-Aid for Scientific Research (C) through Japan Society for the Promotion of Science (JSPS).
R.K. acknowledges support from Grant-in-Aid for JSPS Research Fellow (16J01302).

\bibliographystyle{apj}
\bibliography{bibtex.bib}


\begin{appendix} \label{appen}
\setcounter{figure}{10}
Before clustering analysis in Section \ref{seq_halo}, we calculate the angular correlation functions for all five fields. We separate each sample to luminosity bins, and compare with previous results \citep{ouchi04, lee06, barone14}. Figure \ref{figap_1} shows the angular correlation functions for the GOODS-North and UDS fields. The clustering properties for these two fields are relatively smaller than the values by the previous results. The GOODS-North field has a negative correlation with luminosity. The UDS field has a smaller angular correlation function and there are no signals beyond 100 arcsec. Because of this strange behavior, we does not include these two fields for our analysis. The cause of this weak clustering properties is not clear. The small number of filters used for SED fitting may affect clustering properties. 

\begin{figure}[tbp]
  \centering
      \includegraphics[width= 0.9\linewidth, trim=0 0 0 0, clip]{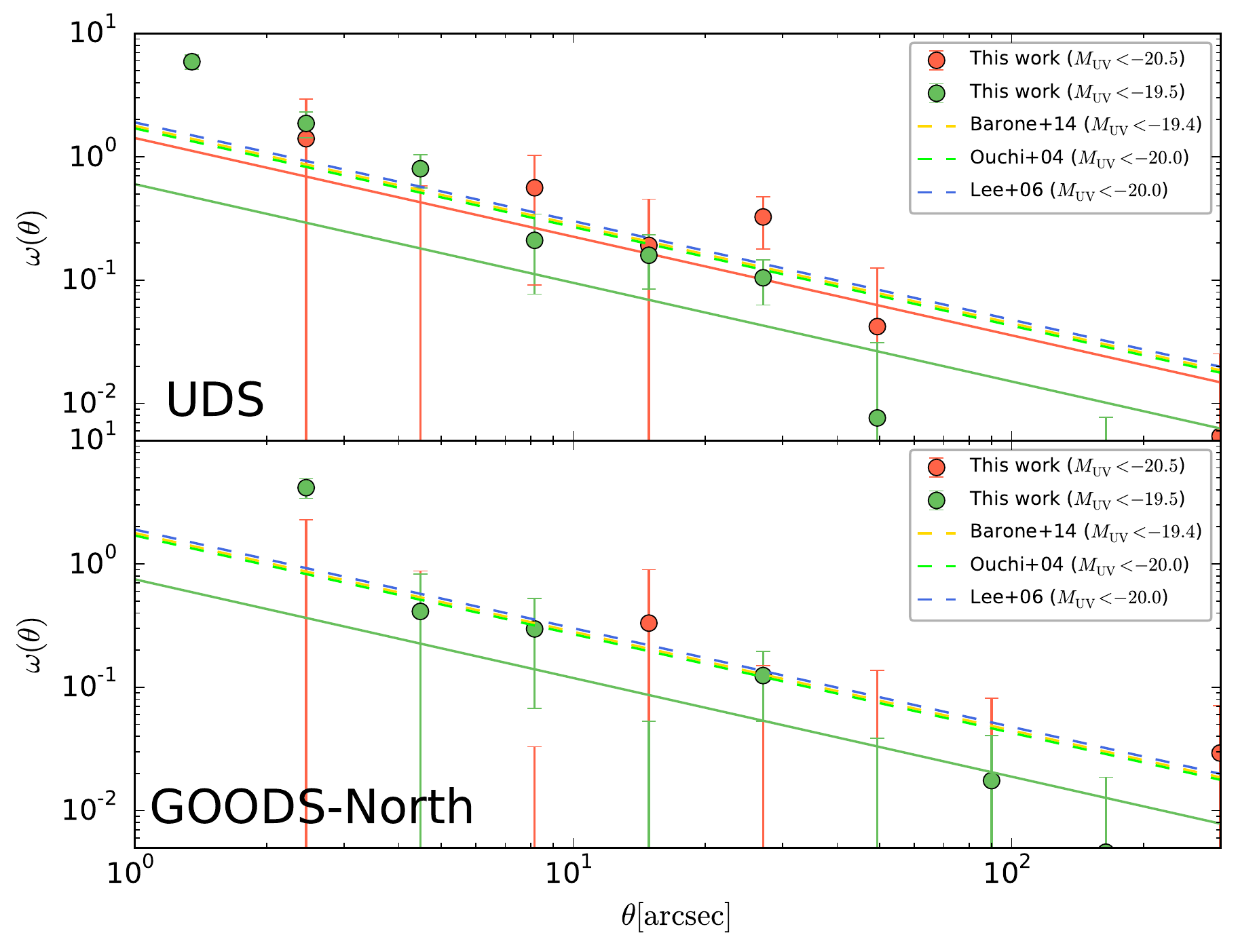}
  \caption{Angular correlation functions in the UDS (top panel) and GOODS-North (bottom) at $z\sim4$ compared with three previous results. The solid red and green lines indicate the best-fit power laws for luminosity bins. The dashed yellow, green, and blue lines indicate the results by \citet{barone14}, \citet{ouchi04}, and \citet{lee06}, respectively.}
  \label{figap_1}
\end{figure}
\end{appendix}

\end{document}